%% file: ANA-HDBS-2019-06-PAPER.tex
\pdfoutput=1
\documentclass[cernpreprint,atlasdraft=true,texlive=2016,UKenglish,texmf]{atlasdoc}

\usepackage{atlaspackage}
\usepackage{atlasbiblatex}
 
\usepackage{atlasphysics}
 
\graphicspath{{logos/}{figures/}}
 
\usepackage{ANA-HDBS-2019-06-PAPER-defs}
\usepackage{multirow}
\usepackage{cancel}

\usepackage[color=orange,]{todonotes}

\AtlasTitle{Search for doubly and singly charged Higgs bosons decaying into vector bosons in multi-lepton final states with the ATLAS detector using proton--proton collisions at $\sqrt{s} = \SI{13}{TeV}$}
 
\AtlasAbstract{
A search for charged Higgs bosons decaying into $W^\pm W^\pm$ or $W^\pm Z$ bosons is performed,
involving experimental signatures with two leptons of the same charge, or three or four leptons with a variety of charge combinations, missing transverse momentum and jets.
A data sample of proton--proton collisions at a centre-of-mass energy of \SI{13}{TeV} recorded with the ATLAS detector at the Large Hadron Collider between 2015 and 2018 is used.
The data correspond to a total integrated luminosity of \SI{139}{fb^{-1}}.
The search is guided by a type-II seesaw model that extends the scalar sector of the Standard Model with a scalar triplet,
leading to a phenomenology that includes doubly and singly charged Higgs bosons.
Two scenarios are explored, corresponding to the pair production of doubly charged $H^{\pm\pm}$ bosons, or the associated production of a doubly charged $H^{\pm\pm}$ boson and a singly charged $H^\pm$ boson.
No significant deviations from the Standard Model predictions are observed.
$H^{\pm\pm}$ bosons are excluded at 95\% confidence level up to \SI{350}{GeV} and \SI{230}{GeV} for the pair and associated production modes, respectively.
}

\author{The ATLAS Collaboration}
 
\AtlasRefCode{HDBS-2019-06}

\AtlasJournalRef{JHEP 06 (2021) 146}
\AtlasDOI{10.1007/JHEP06(2021)146}
\PreprintIdNumber{CERN-EP-2020-240}
 
\hypersetup{pdftitle={ATLAS document},pdfauthor={The ATLAS Collaboration}}

\begin{document}
 
\maketitle

\section{Introduction} 
Experimental signatures with two leptons with the same electric charge (same-charge) or multi-lepton final states are extensively exploited
in searches for physics beyond the Standard Model (BSM physics) at hadron colliders.
In many models, heavy BSM particles may be produced in proton--proton collisions and decay into multiple massive Standard Model (SM) electroweak gauge bosons or top quarks.
Subsequent decays of these gauge bosons into final states with leptons can occur with considerable branching ratios.
These final states are favourable to the search for new phenomena
since the yields predicted within the SM are generally low and the experimental effects are well understood.
At the Large Hadron Collider (LHC)~\cite{Evans:2008zzb}, signatures with two leptons of the same charge, or three or four leptons with a variety of charge combinations have been used by the ATLAS and CMS experiments to explore the landscape of possible SM extensions and their phenomenology~\cite{ATLAS:2020wny,Aad:2020mkp,Aad:2020klt,PhysRevD.87.052002,Aad:2015dha,Aad:2020fpj,Aaboud:2018fvk,Aaboud:2018xpj,ATLAS:2012mn,ATLAS:2014kca,Aad:2011vj,Aad:2019ftg,Aad:2019vvi,Aad:2020zbq,Aad:2015eda,Aad:2015jqa,2020135710,Sirunyan:2018xiv,Chatrchyan:2012fla,Sirunyan:2020ztc}.
Among these proposed extensions, the addition to the SM sector of one weak gauge triplet of scalar fields with a non-zero vacuum expectation value of the neutral component
is a compelling way to account for neutrino masses through the type-II seesaw mechanism~\cite{PhysRevD.22.2227}.
 
Guided by such a model, this paper presents a search for doubly and singly charged Higgs bosons.
In the considered type-II seesaw model, after the electroweak symmetry breaking, there are seven scalar bosons: $H^{\pm\pm}$, $H^{\pm}$, $A^0$ (CP odd), $H^0$ (CP even) and $h^0$ (CP even).
The vacuum expectation value is assumed to be \SI{100}{MeV},
the lightest CP-even Higgs boson is the discovered SM Higgs boson,
and the chosen model parameters are compatible with the measured SM Higgs boson properties and electroweak precision data.
The theoretical and experimental constraints help in
choosing allowed values for the parameters of the model from which the predominant production and decay modes follow.
Only the pair production of doubly charged $H^{\pm\pm}$ bosons via a virtual photon or a $Z$ boson,
and the associated production of a doubly charged $H^{\pm\pm}$ boson and a singly charged $H^{\pm}$ boson via a $W^\pm$ boson are considered.
For the $H^{\pm\pm}$ pair production mode the $H^{\pm\pm}$ boson mass is at least \SI{100}{GeV} lower than the $H^{\pm}$ boson mass,
while for the associated production of $H^{\pm\pm}$ and $H^{\pm}$ bosons the mass difference between the doubly and singly charged Higgs bosons is at most \SI{5}{GeV}.
Signal processes were simulated for different $H^{\pm\pm}$ boson mass hypotheses between $200$ and \SI{600}{GeV}.
The $H^{\pm\pm}$ boson is considered to decay into a $W^{\pm}W^{\pm}$ pair and the $H^{\pm}$ boson into a $\WZ$ pair.
The four gauge-boson final states are identified by the presence of charged leptons (electrons or muons), missing transverse momentum and hadronic jets.
The analysis uses three final states (channels) classified according to the number of leptons: two same-charge (SC) leptons (\ll), three leptons (\lll), and four leptons (\llll).
Electrons and muons from $\tau$-lepton decays are also considered.

An earlier ATLAS analysis using this signature to target the pair production of doubly charged $H^{\pm\pm}$ bosons at
$\sqrt s = \SI{13}{TeV}$ with an integrated luminosity of \SI{36}{fb^{-1}}
placed lower limits of between 200 and \SI{220}{GeV} on the $H^{\pm\pm}$ boson mass~\cite{Aaboud:2018qcu}.
In this paper the search is extended and refined, using a larger data set of $pp$ collisions at $\sqrt s = \SI{13}{TeV}$
recorded by the ATLAS detector corresponding to an integrated luminosity of \SI{139}{fb^{-1}}.
In addition, the production of a $H^{\pm\pm}$ boson in association with a $H^{\pm}$ boson is included, extending the initial search and allowing
a search for doubly and singly charged Higgs bosons in a previously unexplored region of the considered model's parameter space.
Section~\ref{sec:SignalModel} presents the theoretical framework while
Section~\ref{sec:detector} gives an overview of the experimental set-up.
The different Monte Carlo simulations used in the analysis are described in Section~\ref{sec:MCSimulation},
followed by a description of the event reconstruction in Section~\ref{sec:EvtReco}.
Signal regions are built from a set of requirements on the kinematic properties of the selected events, in order to isolate the signature of BSM processes from SM backgrounds, as presented in Section~\ref{sec:EvtSelection}.
Section~\ref{sec:SMandD_bkg} describes the estimation of the backgrounds.
The systematic uncertainties are detailed in Section~\ref{sec:syst_unc}.
Event yields in data are then compared with the estimated contributions from SM processes, and the results are discussed in Section~\ref{sec:result}.
Conclusions are presented in Section~\ref{sec:conclusion}.
 

\section{Signal model} 
\label{sec:SignalModel}
 
A type-II seesaw model, in which
the scalar sector of the SM is extended with a hypercharge $Y = 2$ scalar triplet $\Delta$~\cite{Konetschny:1977bn,Cheng:1980qt,Lazarides:1980nt,Schechter:1980gr,PhysRevD.78.015018,PhysRevD.84.095005}, is studied.
The electroweak symmetry breaking (EWSB) is achieved by requiring
the neutral components of the SM scalar doublet $H$ and scalar triplet $\Delta$ to acquire vacuum expectation values, $v_\mathrm{d}$ and $v_\mathrm{t}$ respectively.
After the EWSB, the mixing between these fields results in seven scalar bosons: $H^{\pm\pm}$, $H^{\pm}$, $A^0$ (CP odd), $H^0$ (CP even) and $h^0$ (CP even).
The $h^0$ boson is the SM Higgs boson,
and a small mixing between the CP-even scalars is naturally obtained in the model.
In addition, the triplet--neutrino Yukawa term provides non-zero neutrino masses proportional to the vacuum expectation value of the triplet $v_\mathrm{t}$.
In this type-II seesaw model, constraints from electroweak precision measurements lead to an upper bound on $v_\mathrm{t}$ of approximately \SI{2.5}{GeV},
which is significantly lower than the electroweak scale and matches the need for small values suggested by the natural association of $v_\mathrm{t}$ with the neutrino masses.
Additional, theoretical constraints are the absence of tachyonic modes, requirements for the vacuum structure and stability of the potential, and unitarity requirements.
A detailed discussion can be found in Ref.~\cite{PhysRevD.84.095005}.
The theoretical and experimental constraints help to
choose allowed values for the parameters of the model from which the predominant production and decay modes follow.
Some type-II seesaw models also extend the weak sector by introducing two scalar triplets and right-handed gauge bosons $W_\mathrm{R}$ (left--right symmetric models).
The model studied in this analysis corresponds to the left-handed triplet discussed in Refs.~\cite{Senjanovic:2016bya,MINKOWSKI1977421,PhysRevLett.44.912,Glashow:1979nm,Mohapatra:1980yp}.

\begin{figure}[t!]\centering
\subfloat[]{
\includegraphics[width=0.35\textwidth]{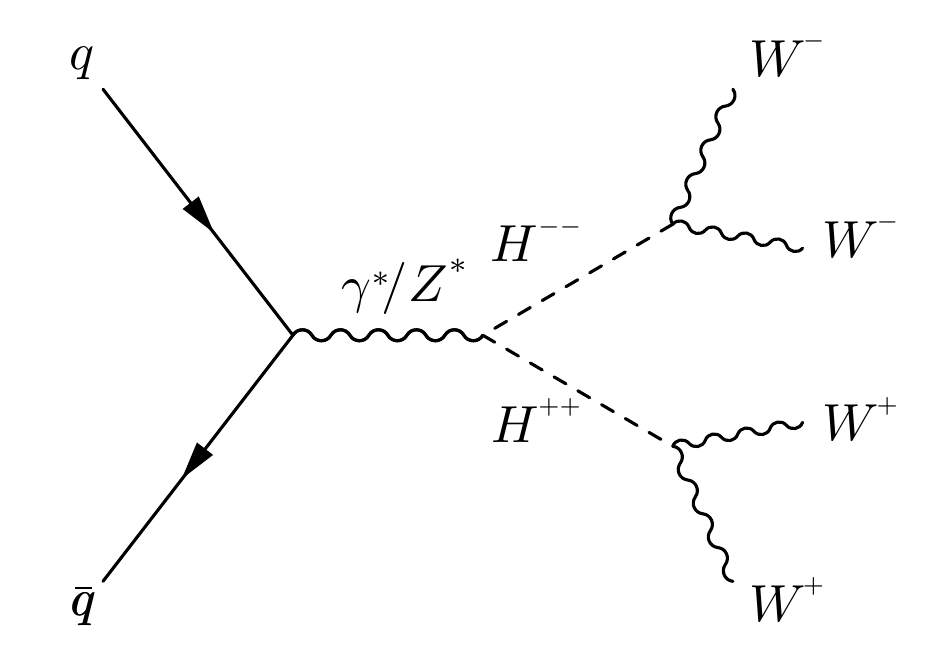}
\label{fig:feyn_pair}
}
\subfloat[]{
\includegraphics[width=0.35\textwidth]{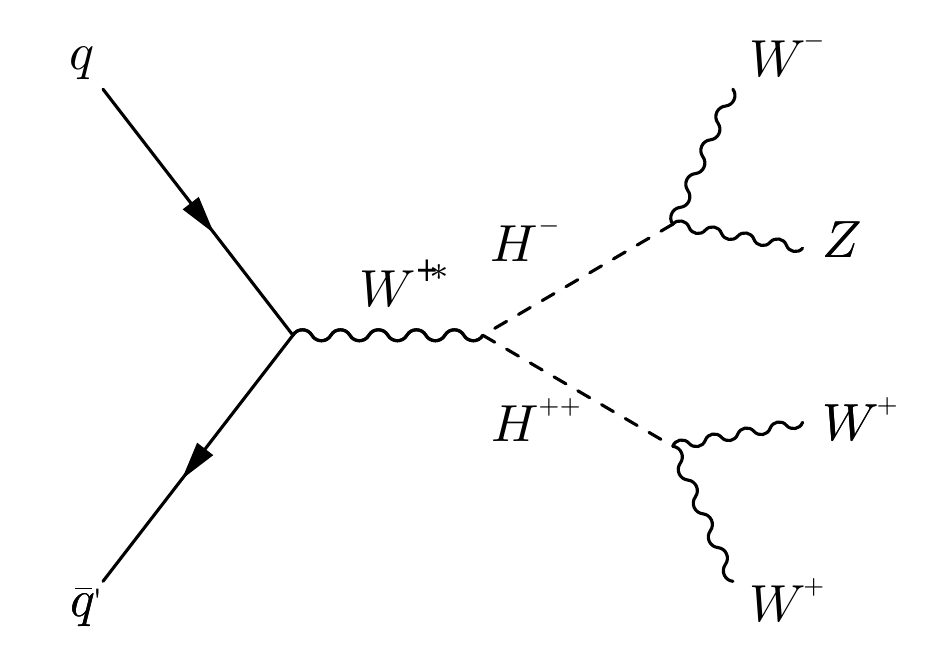}
\label{fig:feyn_asso}
}
\caption{
Feynman diagrams illustrating production and subsequent decay of the $H^{\pm\pm}$ and $H^{\pm}$ bosons:
(a) $H^{\pm\pm}$ pair production and (b) $H^{\pm\pm}$ and $H^{\pm}$ associated production.
}
\end{figure}

Two scenarios with $v_\mathrm{t}$ equal to \SI{100}{MeV} are explored in this paper.
In the first scenario, the mass difference between $H^{\pm\pm}$ and $H^{\pm}$ bosons is $> \SI{100}{GeV}$, the $H^{\pm}$ boson being the heavier one.
Large mass splitting could still be allowed despite the $|m_{H^{\pm}} - m_{H^{\pm\pm}}| < \SI{40}{GeV}$ constraint deduced from the electroweak precision data~\cite{Chun:2012jw} if
radiative corrections reduce the tree-level contribution coming from the triplet vacuum expectation value.
Only the pair production of $H^{\pm\pm}$ bosons via the diagram shown in Figure~\ref{fig:feyn_pair} ($pp\rightarrow \gamma^{*}/Z^{*} \rightarrow H^{\pm\pm}H^{\mp\mp}$) is considered.
The triplet vacuum expectation value is taken to be $v_\mathrm{t}=\SI{100}{MeV}$ such that only the doubly charged $H^{\pm\pm}$ boson decay into
a pair of $W$ bosons with the same charge, $H^{\pm\pm} \rightarrow W^{\pm}W^{\pm}$, is possible~\cite{PhysRevD.78.015018}.
The leptonic decays $H^{\pm\pm} \rightarrow \ell^{\pm}\ell^{\pm}$ are suppressed with increasing $v_\mathrm{t}$~\cite{Kanemura:2014goa,kang:2014jia} and not considered in this paper.

In the second scenario, the mass of the $H^{\pm}$ boson is chosen to be at most \SI{5}{GeV} different from the $H^{\pm\pm}$ boson mass.
The low mass difference ensures that the charged Higgs bosons do not decay into each other with a non-negligible branching fraction,
and complies with the additional $|m_{H^{\pm}} - m_{H^{\pm\pm}}| < \SI{40}{GeV}$ constraint deduced from the electroweak precision data.
In addition, the chosen mass difference between the doubly and singly charged Higgs bosons maximises the signal amplitudes.
Only the associated production of $H^{\pm\pm}$ and $H^{\pm}$ bosons via the diagram shown in Figure~\ref{fig:feyn_asso} ($pp\rightarrow W^{\pm*}\rightarrow H^{\pm\pm}H^{\mp}$) is considered.
The production cross sections for $H^{\pm\pm}$ pair production with the same mass settings as for associated production of $H^{\pm\pm}$ and $H^{\pm}$ bosons can be large.
However, this production mode is exactly the one described in the first scenario and it is not considered in the second scenario.
This choice is motivated by the objective of the search, which is to study a characteristic signature with a benchmark parameter point and report its cross section.
The $H^{\pm\pm}$ boson decays into a pair of $W$ bosons with the same charge, with a branching ratio of 100\%.
Only the bosonic decays of the singly charged bosons ($H^\pm \rightarrow W^\pm Z$) are considered and,
depending on the $H^\pm$ boson mass, the branching ratio varies between 40\% and 60\% (see Table~\ref{Tab:Signal_xSec}).
The branching fraction for the $m_{H^{\pm\pm}} =300$~\GeV mass hypothesis is lower than for the neighbouring points because of the high dependency of this quantity on the mass
of the scalar triplet $\Delta$~\cite{PhysRevD.84.095005}.
Depending on the $H^\pm$ boson mass, other decay modes are $H^{\pm}\rightarrow t \bar{b}$ and $H^{\pm}\rightarrow W^{\pm} h^0$~\cite{PhysRevD.78.015018}.
Studies at Monte Carlo generator level show that after the selection of at least two same-charge leptons or at least three leptons and no $b$-jet (where the $b$-jets are selected with 70\% efficiency),
the contribution from the other possible $H^\pm$ decays is negligible.
Similar conclusions are also reached after examining the results obtained for the various control regions used in the analysis.
The effect on the mass below which a charged Higgs boson is excluded is negligible.

Pair production of $H^{\pm}$ bosons is also possible, albeit with a much smaller production cross section than for $H^{\pm\pm}$ pair production.
Therefore, $H^{\pm}$ pair production is not considered in this paper.
The cross section for single $H^{\pm\pm}$ production via vector-boson fusion ($pp\rightarrow W^{\pm *}W^{\pm *}\rightarrow H^{\pm\pm}$)
is proportional to $v_\mathrm{t}$, and hence negligible.

For the $H^{\pm\pm}$ pair production mode, the mixing  between the CP-even scalars is taken to be $10^{-4}$ and the remaining five couplings in the potential are adjusted
to obtain a given $H^{\pm\pm}$ boson mass hypothesis while requiring $h^{0}$ to have a mass of $125$~\GeV.
Similar settings are also used for the associated production mode.
The branching fraction times cross-section calculation for the pair production of $H^{\pm\pm}$ bosons and the associated production of $H^{\pm\pm}$ and $H^{\pm}$ bosons
is performed for on-shell $W$ and $Z$ bosons, and therefore only the region $m_{H^{\pm\pm}} > 200$~\GeV is considered in the present analysis.
 
Extensive searches for leptonic decays have been performed at various experiments~\cite{Abbiendi:2001cr,Aktas:2006nu,Aaltonen:2011rta,ATLAS:2012hi,Chatrchyan:2012ya,Aaboud:2017qph},
excluding doubly charged $H^{\pm\pm}$ bosons with masses up to about \SI{870}{GeV}.
The CMS Collaboration published results for the $H^{\pm\pm} \rightarrow W^{\pm}W^{\pm}$ decay mode
in the context of single $H^{\pm\pm}$ production through vector-boson fusion with $v_\mathrm{t}$ values of a few tens of \SI{}{GeV}, for a model with two Higgs triplets~\cite{Khachatryan:2014sta,Sirunyan:2017ret,GEORGI1985463}.
In contrast, the $H^{\pm\pm} \rightarrow W^{\pm}W^{\pm}$ decay mode is not often investigated for $v_\mathrm{t}$ values around \SI{100}{MeV}.
In this case, the difference is that single $H^{\pm\pm}$ production is suppressed and only $H^{\pm\pm}$ pair production is sizeable.
A direct search for $H^{\pm\pm}$ pair production with decays to $W^{\pm}W^{\pm}$ pairs has been performed on a smaller data set by the ATLAS Collaboration~\cite{Aaboud:2018qcu}, validating the principle of such an approach.

\section{ATLAS detector} 
\label{sec:detector}
\newcommand{\AtlasCoordFootnote}{
ATLAS uses a right-handed coordinate system with its origin at the nominal interaction point (IP)
in the centre of the detector and the \(z\)-axis along the beam pipe.
The \(x\)-axis points from the IP to the centre of the LHC ring,
and the \(y\)-axis points upwards.
Cylindrical coordinates \((r,\phi)\) are used in the transverse plane,
\(\phi\) being the azimuthal angle around the \(z\)-axis.
The pseudorapidity is defined in terms of the polar angle \(\theta\) as \(\eta \equiv -\ln \tan(\theta/2)\), and is equal to the rapidity $y \equiv 0.5\ln{\left((E+p_{z})/(E-p_{z})\right)}$
in the relativistic limit. Angular distance is measured in units of \(\Delta R \equiv \sqrt{(\Delta y)^{2} + (\Delta\phi)^{2}}\). The magnitude of the momentum in the plane transverse to the beam axis is denoted by \pt.}

The ATLAS experiment~\cite{PERF-2007-01} at the LHC is a multipurpose particle detector
with a forward--backward symmetric cylindrical geometry and nearly \(4\pi\) coverage in
solid angle.\footnote{\AtlasCoordFootnote}
It consists of an inner tracking detector (ID) surrounded by a thin superconducting solenoid
providing a \SI{2}{\tesla} axial magnetic field, electromagnetic (EM) and hadron calorimeters, and a muon spectrometer (MS).
The inner tracking detector covers the pseudorapidity range \(|\eta| < 2.5\).
It consists of silicon pixel, silicon microstrip and transition radiation tracking detectors;
the innermost layer is \SI{33}{mm} from the beamline~\cite{ATLAS-TDR-2010-19,ATLAS-TDR-19-addm,PIX-2018-001}.
Lead/liquid-argon (LAr) sampling calorimeters provide electromagnetic energy measurements
with high granularity.
A steel/scintillator-tile hadron calorimeter covers the central pseudorapidity range (\(|\eta| < 1.7\)).
The endcap and forward regions are instrumented with LAr calorimeters
for EM and hadronic energy measurements up to \(|\eta| = 4.9\).
The muon spectrometer surrounds the calorimeters and is based on
three large air-core toroidal superconducting magnets with eight coils each.
The field integral of the toroids ranges between \num{2.0} and \SI{6.0}{\tesla$\cdot$\metre}
across most of the detector.
The muon spectrometer includes a system of precision tracking chambers and fast detectors for triggering.
A two-level trigger system~\cite{TRIG-2016-01} is used to select events.
The first-level trigger is implemented in hardware and uses a subset of the detector information
to keep the accepted rate below \SI{100}{\kHz}.
This is followed by a software-based trigger that
reduces the accepted event rate to \SI{1}{\kHz} on average
depending on the data-taking conditions.

\section{Monte Carlo event simulation} 
\label{sec:MCSimulation}
\begin{table}[t!]
\begin{center}
\def\arraystretch{1.2}
\caption{
The cross sections for the $H^{\pm\pm}$ pair and the $H^{\pm\pm}$ and $H^{\pm}$ associated production modes at next-to-leading order in QCD~\cite{Muhlleitner:2003me}.
The branching ratios ($\mathcal{B}$) of the charged Higgs bosons to $W^\pm W^\pm$ or $\WZ$ are included in the quoted values.
}
\label{Tab:Signal_xSec}
 
\scriptsize
\resizebox{0.82\textwidth}{!}
{
\begin{tabular}{l||c|c|c|c|c|c|c|c|c}
\hline\hline
 
$m_{H^{\pm\pm}}$ [GeV]                      & \multicolumn{2}{c|}{200} & 300 & 350 & \multicolumn{2}{c|}{400} & \multicolumn{2}{c|}{500} & 600  \\
$m_{H^\pm}$ [GeV]                           & \multicolumn{2}{c|}{400} & 400 & 700 & \multicolumn{2}{c|}{700} & \multicolumn{2}{c|}{700} & 700  \\	\hline
$\mathcal{B}(H^{\pm\pm}\rightarrow W^\pm W^\pm)$ [\%] & \multicolumn{2}{c|}{100} & 100 & 100 & \multicolumn{2}{c|}{100} & \multicolumn{2}{c|}{100} & 100  \\  \hline
Cross section [fb]                          & \multicolumn{2}{c|}{\multirow{2}{*}{$\numRP{81.04}{1}$}} & \multirow{2}{*}{$\numRP{16.48}{1}$} & \multirow{2}{*}{$\numRP{8.70}{1}$} & \multicolumn{2}{c|}{\multirow{2}{*}{$\numRP{4.88}{1}$}} & \multicolumn{2}{c|}{\multirow{2}{*}{$\numRP{1.77}{1}$}} & \multirow{2}{*}{$\numRP{0.73}{1}$}  \\
($H^{\pm\pm}$ pair production)              & \multicolumn{2}{c|}{}    &     &     & \multicolumn{2}{c|}{}    & \multicolumn{2}{c|}{}    &      \\
 
\hline \hline
\addlinespace \addlinespace
\hline \hline
 
$m_{H^{\pm\pm}}$ [GeV]                      & 200   & 220   & \multicolumn{2}{c|}{300}   & 400   & 450   & 500   & 550   & 600   \\
$m_{H^\pm}$ [GeV]                           & 196   & 215   & \multicolumn{2}{c|}{295}   & 395   & 445   & 496   & 545   & 602   \\  \hline
$\mathcal{B}(H^{\pm\pm}\rightarrow W^\pm W^\pm)$ [\%] & 100   & 100   & \multicolumn{2}{c|}{100}   & 100   & 100   & 100   & 100   & 100   \\
$\mathcal{B}(H^\pm \rightarrow W^\pm Z)$ [\%]         & $\numRP{58.76}{1}$ & $\numRP{44.26}{1}$ & \multicolumn{2}{c|}{$\numRP{37.34}{1}$} & $\numRP{44.70}{1}$ & $\numRP{45.92}{1}$ & $\numRP{45.74}{1}$ & $\numRP{48.44}{1}$ & $\numRP{50.76}{1}$ \\  \hline
Cross section [fb]                          & \multirow{2}{*}{$\numRP{88.66}{1}$} & \multirow{2}{*}{$\numRP{44.48}{1}$} & \multicolumn{2}{c|}{\multirow{2}{*}{$\numRP{9.46}{1}$}} & \multirow{2}{*}{$\numRP{2.99}{1}$} & \multirow{2}{*}{$\numRP{1.85}{1}$} & \multirow{2}{*}{$\numRP{1.18}{1}$} & \multirow{2}{*}{$\numRP{0.78}{1}$} & \multirow{2}{*}{$\numRP{0.51}{1}$} \\
($H^{\pm\pm}H^{\mp}$ associated production) &       &       & \multicolumn{2}{c|}{}      &       &       &       &       &       \\
 
\hline\hline
\end{tabular}
}
\end{center}
\end{table}
 
Monte Carlo (MC) event generators were used to simulate the signal and background events produced in the proton--proton collisions.
For the $H^{\pm\pm}$ pair production and the $H^{\pm\pm}$ and $H^{\pm}$ associated production signal processes,
the events at particle level were generated with \mgamc~\cite{Alwall:2014hca} using leading-order (LO) matrix elements (ME) and
\nnpdfloAlt parton distribution functions (PDF)~\cite{Ball:2014uwa}. The events were subsequently showered using
\PYTHIA~{8}~\cite{Sjostrand:2014zea} with a set of tuned parameters called the A14 tune~\cite{ATL-PHYS-PUB-2014-021}.
Signal processes were simulated for different mass hypotheses between $200$ and $600$ GeV, and presented in Table~\ref{Tab:Signal_xSec}.
For the pair and associated production, the cross section decreases rapidly with the charged Higgs boson's mass as shown in Table~\ref{Tab:Signal_xSec}.
The event samples were normalised using calculations at next-to-leading order (NLO) in perturbative QCD~\cite{Muhlleitner:2003me}.
The NLO $K$-factor increases the expected event yields by a factor of 1.25, independently of the mass of the charged Higgs bosons.

A summary of the MC event generators used to simulate the background
is presented in Table~\ref{Tab:samples} and further details are given below.
The notation $V$ is used to refer to an electroweak gauge boson $W$ or $Z/\gamma^*$.

\begin{table}[t!]
\begin{center}
\def\arraystretch{1.2}
\caption{
Summary of the event generators, parton shower models, and PDF sets used for the simulation of the background event samples.
The notation $V$ is used to refer to an electroweak gauge boson $W$ or $Z/\gamma^*$.
In the final column, `default' refers the to default parameter set provided with the event generator.
}
\label{Tab:samples}
 
\scriptsize
\resizebox{\textwidth}{!}
{
\begin{tabular}{l||c c c c c}
\hline\hline
Process & Generator & ME accuracy & PDF & Parton shower and hadronisation & Parameter set  \\
\hline\hline

$VV$, $V\gamma$ & \sherpa & NLO (0-1j) + LO (2-3j) & \texttt{NNPDF3.0nnlo} & \sherpa{} & default \\
$VV$-EW~jj & \sherpa & LO & \texttt{NNPDF3.0nnlo} & \sherpa{} & default \\
 
$VVV$ & \sherpa & NLO(0j) + LO (1-2j) & \texttt{NNPDF3.0nnlo} & \sherpa & default \\
 
$V$+jets & \sherpa & NLO (0-2j) + LO (3-4j) & \texttt{NNPDF3.0nnlo} & \sherpa{} & default \\
 
$VH$ & \PYTHIA~{8} & LO & \texttt{NNPDF2.3lo} & \PYTHIA~{8} & A14 \\

\ttH\ & \POWHEGBOX~{v2} & NLO & \texttt{NNPDF3.0nlo} & \PYTHIA~{8} & A14 \\
 
\ttV, \tWZ, \tZ\ & \mgamc & NLO & \texttt{NNPDF3.0nlo} & \PYTHIA~{8} & A14 \\

$t\bar{t}$, \tW\ & \POWHEGBOX~{v2} & NLO &\texttt{ NNPDF3.0nnlo} & \PYTHIA~{8} & A14 \\
 
\tttt, \ttt\, \ttWW, \ttWZ\ & \mgamc & NLO & \texttt{NNPDF3.1nlo} & \PYTHIA~{8} & A14 \\

\hline\hline
\end{tabular}
}
\end{center}
\end{table}
 
Event samples of diboson ($VV$) and $V\gamma$ final states were simulated with the
\sherpa~v2.2.1, v2.2.2 or v2.2.4~\cite{Bothmann:2019yzt} generator depending on the process, including off-shell effects and Higgs boson contributions, where appropriate.
The $VV$ and $V\gamma$ processes were simulated using matrix elements at NLO accuracy in QCD for up to one additional parton and at LO accuracy for up to three additional parton emissions.
The electroweak production of dibosons in association with two jets ($VV$-EW~jj) was generated using LO accuracy matrix elements.
The matrix element calculations were matched and merged with the \sherpa parton shower based on Catani--Seymour dipole
factorisation~\cite{Gleisberg:2008fv,Schumann:2007mg} using the MEPS@NLO prescription~\cite{Hoeche:2011fd,Hoeche:2012yf,Catani:2001cc,Hoeche:2009rj}. The virtual QCD corrections were provided by
the \openloops\ library~\cite{Cascioli:2011va,Denner:2016kdg}.
The production of triboson ($VVV$) events was simulated with the \sherpa~v2.2.2 generator using factorised gauge-boson decays.
All diboson and triboson processes were generated using the \nnpdfnnlo PDF set~\cite{Ball:2014uwa}, along with the dedicated set of tuned parton-shower parameters developed by the~\sherpa authors.
The cross sections from the event generator were used for the normalisation.
 
The production of an electroweak gauge boson or virtual photon in association with jets ($V$+jets) was simulated with the \sherpa~v2.2.1
generator using NLO matrix elements for up to two partons, and LO matrix elements
for up to four partons calculated with the Comix~\cite{Gleisberg:2008fv} and \openloops\ libraries.
The event samples were generated using the \texttt{NNPDF3.0nnlo} PDF set, along with the dedicated set of tuned parton-shower parameters
developed by the~\sherpa authors. The $V$+jets event samples were normalised to a NNLO prediction.
 
Higgs boson production in association with a vector boson was simulated at LO with \PYTHIA~{8.186} and \evtgen~\cite{EvtGen} using the A14 tune, along with the \texttt{NNPDF2.3lo} PDF set.
The Monte Carlo prediction was normalised to cross sections calculated at NNLO in QCD with NLO electroweak corrections for $q\bar{q}/qg \to VH$, and at next-to-leading-logarithm
accuracy in QCD with NLO EW corrections for $gg \to ZH$~\cite{Ciccolini:2003jy,Brein:2003wg,Brein:2011vx,Altenkamp:2012sx,Denner:2014cla,Brein:2012ne,Harlander:2014wda}.
The production of \ttH\ events was modelled using the \POWHEGBOX~{v2}~\cite{Frixione:2007nw,Nason:2004rx,Frixione:2007vw,Alioli:2010xd,Hartanto:2015uka} generator at NLO with the \nnpdfnlo PDF set~\cite{Ball:2014uwa}.
The events were interfaced to \PYTHIA~{8.230} using the A14 tune and the \nnpdftwo PDF set~\cite{Ball:2014uwa}. The decays of bottom and charm hadrons were performed by \evtgen~v1.6.0.
The production of \ttV, \tWZ\ and \tZ\ events was modelled using the \mgamc~v2.3.3 generator at NLO with the \nnpdfnlo\ PDF.
The events were interfaced to \PYTHIA~{8.210} (\PYTHIA~{8.212} for \tWZ) using the A14 tune and the \nnpdftwo\ PDF set. The decays of bottom and charm hadrons were simulated using the \evtgen\ v1.2.0 program.
The simulated \ttH, \ttV, \tWZ\ and \tZ\ events were normalised to cross-section calculations at NLO accuracy in QCD. In addition, NLO EW corrections were included for \ttH, \ttV\ and \tWZ.
 
Backgrounds from top-quark pair production ($t\bar{t}$) and $tW$ production
were estimated at NLO accuracy in QCD using the hvq program~\cite{Frixione:2007nw} in \POWHEGBOX~{v2}. The event samples were generated using the A14 tune and \texttt{NNPDF3.0nnlo} PDF set.
The interference between $t\bar{t}$ and $tW$ production is neglected as it has a negligible impact on the analysis.
The \ttbar\ event sample was normalised to the cross-section prediction at NNLO in QCD including the resummation of next-to-next-to-leading logarithmic (NNLL) soft-gluon terms calculated using
\textsc{Top++2.0}~\cite{Beneke:2011mq,Cacciari:2011hy,Baernreuther:2012ws,Czakon:2012zr,Czakon:2012pz,Czakon:2013goa,Czakon:2011xx}.
The inclusive $tW$ cross section was corrected to the theory prediction calculated at NLO in QCD with NNLL soft gluon corrections~\cite{Kidonakis:2010ux,Kidonakis:2013zqa}.
 
The production of \tttt, \ttt, \ttWW\ and \ttWZ\ events was modelled using the \mgamc~v2.3.3
generator at NLO with the \nnpdfonenlo PDF~\cite{Ball:2014uwa}.
The events were interfaced with \PYTHIA~{8.230} using the A14 tune and the \nnpdftwo PDF set.
The \tttt, \ttWW\ and \ttWZ\ contributions were normalised to theoretical cross sections at NLO in QCD including electroweak corrections~\cite{Frederix:2017wme}.
The \ttt\ production was normalised to a cross section calculated to LO in QCD.
 
Backgrounds such as $pp \rightarrow W^+ W^+ W^- W^-$ or $pp \rightarrow W^\pm W^\pm Z W^\mp$ with \ll, \lll or \llll final states, which have the same signature as the considered signal,
have very small production cross sections times branching ratio and their contribution in the preselection, control and signal regions was found to be negligible.
 
The signal and background events were passed through the $\GEANT4$~\cite{Agostinelli2003250} simulation of the ATLAS detector~\cite{Aad:2010ah} and reconstructed using the same algorithms as are used for
the data. The effect of multiple proton--proton interactions in the same or nearby bunch crossings (pile-up) is accounted for using inelastic proton--proton interactions generated by \PYTHIA~{8}~\cite{Sjostrand:2007gs}, with the A3 tune~\cite{ATL-PHYS-PUB-2016-017} and the \nnpdftwo PDF set~\cite{Ball:2012cx}. These inelastic proton--proton interactions were added to the signal and background event samples and weighted such that the distribution of the average number of proton--proton interactions in simulation matches that observed in the data.
 

\section{Event reconstruction} 
\label{sec:EvtReco}
The analysis is performed in $pp$ collision data recorded by the ATLAS detector between 2015 and 2018.
In this period, the LHC delivered colliding beams with a peak instantaneous luminosity up to $L=\SI{2.1e34}{cm^{-2}s^{-1}}$, achieved in 2018, and an average number of $pp$ interactions per bunch crossing of \num{34}.
After requirements on the stability of the beams, the operational status of all ATLAS detector components, and the quality of the recorded data,
the total integrated luminosity of the data set corresponds to \SI{139}{fb^{-1}}~\cite{DAPR-2018-01}.
 
Proton--proton interaction vertices are reconstructed from charged-particle tracks with ${\pt>\SI{500}{MeV}}$~\cite{ATL-PHYS-PUB-2015-026,PERF-2015-01} in the ID.
The presence of at least one such vertex with a minimum of two associated tracks is required,
and the vertex with the largest sum of \pTsq\ of associated tracks is chosen as the primary vertex.

The anti-$k_t$ algorithm~\cite{Cacciari:2008gp,Fastjet} with radius parameter $R=0.4$ is used to reconstruct jets up to $\abseta=4.9$.
It uses as inputs particle-flow objects, combining tracking and calorimetric information~\cite{PERF-2015-09}.
The jets are calibrated as described in Ref.~\cite{JETM-2018-05}.
Only jets with $\pt>\SI{20}{GeV}$ and $\abseta<2.5$ are considered further.
Events are removed if they contain jets induced by calorimeter noise or non-collision background,
according to criteria similar to those described in Ref.~\cite{ATLAS-CONF-2015-029}.
Jets with $\pt<\SI{60}{GeV}$ and $\abseta<2.4$ from pile-up interactions are suppressed using a jet-vertex tagging discriminant~\cite{PERF-2014-03}.
 
Jets containing $b$-flavored hadrons are identified in the region $\abseta<2.5$ with the DL1r \btag algorithm based on a recurrent neural network~\cite{FTAG-2018-01,ATL-PHYS-PUB-2017-013}.
It makes use of the impact parameters of tracks associated with the jet,
the position of reconstructed secondary vertices and their compatibility with the decay chains of such hadrons.
The $b$-tagging average efficiency is 70\%, as measured in $t\bar t$ events.

Electron candidates are reconstructed as tracks in the ID matched to energy clusters in the EM calorimeter, within $|\eta| <  2.47$~\cite{EGAM-2018-01}.
Only electrons with $\pt > \SI{10}{GeV}$ and not in the transition region between the barrel and endcap calorimeters ($1.37  < |\eta| <  1.52$) are considered.
The electron identification is based on a multivariate likelihood-based discriminant that uses the shower shapes in the EM calorimeter and the associated track properties measured in the ID.
The electron candidates must satisfy the `Loose' identification criteria described in Ref.~\cite{EGAM-2018-01}.
Signal electrons are required to satisfy the `Tight' identification~\cite{EGAM-2018-01} for better rejection of non-prompt electrons.
The electron identification efficiency varies with increasing $\pt$ in $Z \to ee$ events, from 65\% at $\pt = \SI{10}{GeV}$ to 88\% at $\pt = \SI{100}{GeV}$ for the {Tight} operating point, and from 85\% at
$\pt = \SI{20}{GeV}$ to 95\% at $\pt = \SI{100}{GeV}$ for the {Loose} operating point.
 
The longitudinal impact parameter of the electron track, $z_0$, is required to satisfy $|z_{0}\sin\theta| < \SI{0.5}{mm}$, where $\theta$ is the polar angle of the track.
The transverse impact parameter divided by its uncertainty, $|d_{0}|/\sigma(d_{0})$, is required to be less than five.
For all signal electrons there must be no association with a vertex from a reconstructed photon conversion~\cite{EGAM-2018-01} in the detector material. 
To further reduce the photon conversion background in the \ll channel, additional requirements are applied to the signal electrons~\cite{Aad:2020klt}:
i) the candidate must not have a reconstructed displaced vertex with radius $r > \SI{20}{mm}$ whose reconstruction uses the track associated with the electron,
ii) the invariant mass of the system formed by the track associated with the electron and the closest track at the primary or a conversion vertex
is required to be larger than $\SI{100}{MeV}$;
this selection is referred to as the photon conversion veto. 
 
For the signal electrons, the identification criteria are complemented by an isolation requirement,
which is based on the energy in a cone around the electron candidate calculated using either reconstructed tracks or energy clusters.
Electrons with wrongly reconstructed charge (charge-flip)
are suppressed using a boosted decision tree discriminant exploiting additional tracks in the vicinity of the electron and track-to-cluster matching variables~\cite{EGAM-2018-01}.

Muon candidates are reconstructed in the region $\abseta<2.5$ from MS tracks matching ID tracks.
The analysis only considers muons with $\pt>\SI{10}{GeV}$ satisfying the `Medium' quality requirements defined in Ref.~\cite{PERF-2015-10}.
The muon reconstruction efficiency is approximately 98\% in $Z \to \mu\mu$ events.
The longitudinal impact parameter of the muon track must satisfy $|z_0\sin\theta|<\SI{0.5}{mm}$ and
the transverse impact parameter must satisfy $|d_0|/\sigma(d_0) <3$.
For signal muons the candidate must satisfy calorimeter- and track-based isolation requirements.

Non-prompt electrons and muons from the decays of $b$- and $c$-flavored hadrons are further rejected
using a boosted decision tree discriminant based on isolation and secondary vertex information,
referred to as the non-prompt-lepton veto~\cite{Aaboud:2017jvq}.

To avoid cases where the detector response to a single physical object is reconstructed as two different final-state objects,
e.g.~an electron reconstructed as both an electron and a jet, several steps are followed to remove such overlaps, as described in Ref.~\cite{Aad:2019ftg}.
The overlap removal procedure is performed using candidate leptons.

\begin{table}[t!]
\begin{center}
\def\arraystretch{1.2}
\caption{
The requirements applied to define the categories of candidate leptons: loose (L), loose and minimally isolated (L$^*$) and tight (T) leptons.
The overlap removal procedure is not applied for the candidate leptons.
In the \ll channel, the photon conversion veto is required in addition to the loose and tight criteria.
}
\label{tab:lep_sel}
 
\scriptsize
\resizebox{0.88\textwidth}{!}
{
\begin{tabular}{l||c|c|c|c||c|c|c|c}
\hline\hline
& \multicolumn{4}{c||}{Electrons} & \multicolumn{4}{c}{Muons} \\
\hline
& Candidate & L & L$^{*}$ & T &     Candidate & L & L$^{*}$ & T \\
\hline
\hline
$|z_{0}\sin\theta|$ & \multicolumn{4}{c||}{$< \SI{0.5}{mm}$} & \multicolumn{4}{c}{$< \SI{0.5}{mm}$} \\ \hline
$|d_{0}|/\sigma(d_{0})$& \multicolumn{4}{c||}{$< 5$} & \multicolumn{4}{c}{$< 3$} \\ \hline
Identification & \multicolumn{3}{c|}{{Loose}} & {Tight} & \multicolumn{4}{c}{{Medium}} \\\hline
Isolation      & \multicolumn{2}{c|}{No} &  Loose & Yes &  \multicolumn{2}{c|}{No} & FixedCutLoose & Yes \\ \hline
Non-prompt-lepton veto & \multicolumn{3}{c|}{No} & Yes  & \multicolumn{3}{c|}{No} & Yes \\ \hline
Electron charge-flip veto & No & \multicolumn{3}{c||}{Yes}  & \multicolumn{4}{c}{N/A}\\
 
\hline\hline
\end{tabular}
}
\end{center}
\end{table}
The different lepton selections used in the analysis are summarised in Table~\ref{tab:lep_sel}. Three types of signal lepton requirements are used for both the
electrons and muons: `tight', `loose', and `loose and minimally isolated'.
The tight leptons, and the loose and minimally isolated leptons, are a subset of the loose leptons.
The photon conversion veto is applied on top of the tight and loose electron selection requirements, only in the \ll channel.
The Loose and FixedCutLoose isolation criteria applied for the loose and minimally isolated electrons and muons are described in Refs.~\cite{EGAM-2018-01} and~\cite{PERF-2015-10}.

The missing transverse momentum, with magnitude \met, is defined
as the negative vector sum of the transverse momenta of all identified objects
(muon candidates, electron candidates and jets) and an additional soft term~\cite{PERF-2016-07,ATLAS-CONF-2018-023}.
The soft term is constructed from all tracks that are matched to the primary vertex and are not
associated with any other objects.


\section{Event selection} 
\label{sec:EvtSelection}
\begin{table}[t!]
\begin{center}
\def\arraystretch{1.2}
\caption{
The preselection criteria for the \ll, \lll and \llll analysis channels.
The leptons are required to pass the loose (L), loose and minimally isolated (L$^*$) or tight (T) requirements.
The leptons are ordered by decreasing $\pt$ ($\ell_1,\ell_2,...$) in the \ll and \llll channels,
while for the \lll channel $\ell_1$ and $\ell_2$ denote the two same-charge leptons and $\ell_0$ denotes the lepton with a charge opposite to the total lepton charge. $Q_\ell$ denotes the charge of each lepton.
SFOC refers to same-flavour opposite-charge lepton pairs.
The symbol ``--'' means no requirement is made.
The equal sign $(=)$ is used to emphasise that the selection criterion has to be exactly the given number.
}
\label{Tab:preselection}
 
\scriptsize
\resizebox{0.88\textwidth}{!}
{
\begin{tabular}{l||c|c|c}
\hline\hline
Selection criteria & \ll & \lll & \llll \\
\hline\hline
\multicolumn{4}{c}{At least one offline tight lepton with $\pt^{\ell}>30$~\GeV that triggered the event} \\
\hline
$N_\ell$ (type L)   & =2 & =3 & =4\\
$N_\ell$ (type L$^*$) & -- & -- & =4\\
$N_\ell$ (type T)   & =2 & $\geq$2 ($\ell_{1,2}$) & $\geq$1 \\
$|\sum Q_\ell|$ & =2 & =1& $\neq$4 \\
Lepton $\pt$ & $\pt^{\ell_1,\ell_2}>30, 20$~\GeV &    $\pt^{\ell_0,\ell_1,\ell_2}>10, 20, 20$~\GeV  &  $\pt^{\ell_1,\ell_2,\ell_3,\ell_4}>10$~\GeV \\
\hline
 
\hline
\MET &  $>70$~\GeV & $>30$~\GeV &  $>30$~\GeV \\
\hline
\nj  &  $\ge 3$ & $\ge 2$ &  -- \\
\hline
$N_{b\textrm{-jets}}$  & \multicolumn{3}{c}{ =0 } \\
\hline
Low SFOC $m_{\ell\ell}$ veto & --  & \multicolumn{2}{c}{$m_{\ell\ell}^\text{oc}>15~\GeV$}  \\
\hline
$Z$ boson decay veto &  $|m_{ee}^\text{sc}- m_{Z}| > 10$~\GeV  & \multicolumn{2}{c}{$|m_{\ell\ell}^\text{oc}- m_{Z}| > 10$~\GeV} \\
 
\hline\hline
\end{tabular}
}
\end{center}
\end{table}
 
 
Candidate events are selected for read-out using lepton triggers that require the one electron or muon to satisfy
identification criteria similar to those used in the offline reconstruction, isolation criteria, and a transverse momentum
requirement of $\pt > \SI{26}{GeV}$~\cite{TRIG-2018-05,TRIG-2018-01-UpDated}.
With increasing \pt the requirements on identification and isolation become less stringent.
All events must contain at least one offline tight lepton with $\pt^{\ell}>30$~\GeV that triggered the event.
The event selection proceeds in two steps: the preselection and the signal regions (SRs) selection.

The preselection is defined in Table~\ref{Tab:preselection}; the three channels (\ll, \lll and \llll) are defined to be mutually exclusive.
Events are selected only if the absolute value of the sum of charges of the leptons is two, one, and two or zero for the \ll, \lll and \llll channels, respectively.
In the \ll channel, the second-highest-\pt lepton is required to have $\pt > 20$~\GeV and both leptons are required to be tight.
Similarly in the \lll channel, each lepton in the pair of leptons of the same charge is required to have $\pt > 20$~\GeV and be tight.
 
Further preselection requirements are based on \MET,  the jet multiplicity \nj and the number of $b$-jets, $N_{b\textrm{-jet}}$.
The lowest \met value is \SI{30}{GeV} in the \lll and \llll channels, and \SI{70}{GeV} in the \ll channel;
this selection helps to reduce the non-prompt lepton, electron charge-flip and $VV$ backgrounds.
In the \ll ($\lll$) channel only events with at least three (two) jets are considered.
The background from top-quark production is highly reduced by requiring zero $b$-jets in the event.
In order to reduce the background from Drell--Yan processes and neutral mesons, the invariant mass of same-flavour opposite-charge
lepton pairs is required to be greater than \SI{15}{GeV} for the \lll and \llll channels.
The invariant mass of the same-flavour opposite-charge lepton pair must differ from the nominal $Z$-boson mass by $10$~\GeV.
For the \ll channel, the $Z$-boson invariant mass veto is also applied to $e^\pm e^\pm$ events, in order to reduce the contributions originating from electron charge misidentification.

\begin{table}[t!]
\begin{center}
\def\arraystretch{1.2}
\caption{
Definition of the signal regions optimised for the study of different $H^{\pm\pm}$ and $H^\pm$ mass hypotheses.
The selection is applied on top of the preselection defined in Table~\ref{Tab:preselection}.
For the $H^{\pm\pm}$ pair production mode, the $m_{H^{\pm\pm}}=300$~\GeV signal regions are also used for $m_{H^{\pm\pm}}=350$~\GeV.
For the $H^{\pm\pm}$ and $H^{\pm}$ associated production mode, the $m_{H^{\pm\pm}}=200$~\GeV, \SI{400}{GeV} and \SI{500}{GeV} signal regions are also used for
$m_{H^{\pm\pm}}=220$~\GeV, \SI{450}{GeV} and \SI{550}{GeV}, respectively.
The variables used are described in Section~\ref{sec:EvtSelection}.
}
\label{Tab:SRselection}
 
\scriptsize
\resizebox{0.89\textwidth}{!}
{
\begin{tabular}{l||c|c|c|c}
\hline\hline
Charged Higgs  & \multirow{2}{*}{$m_{H^{\pm\pm}} = \SI{200}{GeV}$} & \multirow{2}{*}{$m_{H^{\pm\pm}} = \SI{300}{GeV}$} & \multirow{2}{*}{$m_{H^{\pm\pm}} = \SI{400}{GeV}$} &  \multirow{2}{*}{$m_{H^{\pm\pm}} = \SI{500}{GeV}$} \\
boson mass     & & & & \\
 
\hline\hline
Selection criteria & \multicolumn{3}{l}{ \ll channel} \\
\hline\hline
 
$m_\mathrm{jets}$ [\SI{}{GeV}]                 & [100, 450] &  [100, 500] &  [300, 700]  & [400, 1000]  \\
$S$                                      & $<$0.3     &  $<$0.6     &  $<$0.6      &  $<$0.9      \\
$\Delta R_{\ell^{\pm}\ell^{\pm}}$        & $<$1.9     &  $<$2.1     &  $<$2.2      &  $<$2.4      \\
$\Delta\phi_{\ell\ell,\met}$             & $<$0.7     &  $<$0.9     &  $<$1.0      &  $<$1.0      \\
$m_{x\ell}$ [\SI{}{GeV}]                       & [40, 150]  &  [90, 240]  &  [130, 340]  &  [130, 400]  \\
\met [\SI{}{GeV}]                              & $>$100     &  $>$130     &  $>$170      &  $>$200      \\

\hline\hline
Selection criteria & \multicolumn{3}{l}{ \lll channel } \\
\hline\hline
 
$\Delta R_{\ell^{\pm}\ell^{\pm}}$        &  [0.2, 1.7]  &  [0.0, 2.1]  &  [0.2, 2.5]  &  [0.3, 2.8]  \\
$m_{x\ell}$ [\SI{}{GeV}]                       &  $>$160~~      &  $>$190~~      &  $>$240~~      &  $>$310~~      \\
\met [\SI{}{GeV}]                              &  $>$30       &  $>$55       &  $>$80       &  $>$90       \\
$\Delta R_{\ell\mathrm{jet}}$            &  [0.1, 1.5]  &  [0.1, 2.0]  &  [0.1,2.3]   &  [0.5, 2.3]  \\
$\pt^{\text{leading jet}}$ [\SI{}{GeV}]        &  $>$40       &  $>$70       &  $>$100~~      &  $>$95       \\

\hline\hline
Selection criteria & \multicolumn{3}{l}{ \llll channel } \\
\hline\hline
 
$m_{x\ell}$ [\SI{}{GeV}]                                 &  $>$230~~      &  $>$270~~      &  $>$360~~      &  $>$440~~      \\
\met [\SI{}{GeV}]                                        &  $>$60       &  $>$60       &  $>$60       &  $>$60       \\
$\pt^{\ell_1}$ [\SI{}{GeV}]                              &  $>$65       &  $>$80       &  $>$110~~      &  $>$130~~      \\
$\Delta R^\text{min}_{\ell^{\pm}\ell^{\pm}}$       &  [0.2, 1.2]  &  [0.2, 2.0]  &  [0.5, 2.4]  &  [0.6, 2.4]  \\
$\Delta R^\text{max}_{\ell^{\pm}\ell^{\pm}}$       &  [0.3, 2.0]  &  [0.5, 2.6]  &  [0.4, 3.1]  &  [0.6, 3.1]  \\

\hline\hline
\end{tabular}
}
\end{center}
\end{table}

In addition to \MET the following variables are used to define SRs:
\begin{itemize}
\item The invariant mass of all selected leptons in the event, $m_{x\ell}$, where $x$ can be 2, 3 or 4 corresponding to the \ll, \lll or \llll channels.
\item The invariant mass of all jets in the event, $m_\mathrm{jets}$. When there are more than four jets in the event, only the leading four jets are used. This variable is only used for the \ll channel.
\item The distance in $\eta$--$\phi$ between two same-charge leptons, $\Delta R_{\ell^\pm\ell^\pm}$.
It is used for the \ll and \lll channels. In the \llll channel, two such variables can be calculated per event, $\Delta R^\mathrm{min}_{\ell^\pm\ell^\pm}$ and $\Delta R^\mathrm{max}_{\ell^\pm\ell^\pm}$, denoting the minimum and maximum values, respectively.
\item The transverse momentum of the highest-$\pt$ jet, $\pt^{\text{leading jet}}$. This variable is used in the \lll channel.
\item The transverse momentum of the highest-$\pt$ lepton, $\pt^{\ell_1}$. This variable is used in the \llll channel.
\item The azimuthal distance between the dilepton system and $\MET$, $\Delta \phi_{\ell\ell,\MET}$. It is only used in the \ll channel.
\item The smallest distance in $\eta$--$\phi$ between any lepton and its closest jet, $\Delta R_{\ell\mathrm{jet}}$. This variable is used in the \lll channel.
\item The variable $S$, used for the \ll channel to describe the event topology in the transverse plane, and defined using the spread of the $\phi$ angles of the leptons, $\MET$, and jets as follows:
\begin{equation}
S = \frac{{\cal{R} }(\phi_{\ell_1},\phi_{\ell_2},\phi_{\MET})\cdot{\cal{R} }(\phi_{j1},\phi_{j2},\cdots  )}{{\cal{R} }(\phi_{\ell_1,},\phi_{\ell_2},\phi_{\MET},\phi_{j1},\phi_{j2},\cdots) }, \nonumber
\end{equation}
where the $\cal{R} $ is the root mean square that quantifies the spread, ${\cal{R}} (\phi_1, \cdots, \phi_n ) = \sqrt{\frac{1}{n} \sum\limits_{i=1}^{n} (\phi_i - \overline{\phi} )^2}$.
\end{itemize}
These variables are found to discriminate between the signal and the background.
They exploit both the boosted decay topology of the charged Higgs bosons and the high energy of the decay products.

Signal regions are defined for each channel, as summarised in Table~\ref{Tab:SRselection}.
The selection criteria defining the signal regions result from a scan of the multidimensional parameter space of the discriminating variables mentioned above.
The SRs were designed by optimising the sensitivity for the $H^{\pm\pm}$ pair production mode, using the $m_{H^{\pm\pm}} = 200$, 300, 400 and \SI{500}{GeV} mass hypotheses.
The same SRs are used to study the $H^{\pm}$ associated production mode;
even though this approach is not optimal, it is still preferred instead of significantly increasing the number of signal regions.
Additional mass hypotheses are used in the signal region defined for the lower mass hypotheses, since the signal discrimination power does not vary significantly in this regime.
In particular, for the $H^{\pm\pm}$ pair production mode the $m_{H^{\pm\pm}}=300$~\GeV signal regions are also used for $m_{H^{\pm\pm}}=350$~\GeV.
For the $H^{\pm\pm}$ and $H^{\pm}$ associated production mode, the $m_{H^{\pm\pm}}=200$~\GeV, \SI{400}{GeV} and \SI{500}{GeV} signal regions are also used for
$m_{H^{\pm\pm}}=220$~\GeV, \SI{450}{GeV} and \SI{550}{GeV}, respectively.
The SRs defined for the \ll channel are further divided into $ee$, $e\mu$ and $\mu\mu$ final states.
Events in the \lll SRs are separated into two categories according to whether or not a same-flavour opposite-charge lepton pair exists in the event.
This procedure further improves the expected significance by exploring differences in background composition
and lepton-flavour composition between signal and background.
The number of events observed in data is shown together with the expected signal and estimated background yields in Section~\ref{sec:result}.

\section{Background estimation} 
\label{sec:SMandD_bkg}
The background sources can be divided into two main categories.
One category is populated by the SM events which contain only reconstructed charged (prompt) leptons originating from leptonic decays of $W$ and $Z$ bosons.
The second category is formed by non-prompt leptons and charge-flip electrons.
The non-prompt-lepton category refers to the leptons that originate from decays of $b$- and $c$-hadrons,
or single pions that mimic electron signatures.
The electrons from hadron decays into photons which convert into pairs of electrons in the beam pipe or detector material also enter this category.
Lepton candidates reconstructed from these different sources
share the properties of being less isolated, having larger impact parameters relative to the primary vertex and being less likely to satisfy the lepton identification criteria.

The backgrounds from SM processes with prompt leptons are estimated with MC simulations,
except for background from  $WZ$ production, for which the normalisation is corrected using data in a dedicated control region.
Background from non-prompt leptons and electron charge-flip are estimated using data-based methods.
Background from $V\gamma$ production can contribute in the SRs if the photon converts to an electron--positron pair, and is estimated using MC simulations.
Background from $WW$ production is estimated from simulation if the two $W$ bosons have the same electric charge
and from data if the two $W$ bosons have opposite electric charge, since the latter only contributes through electron charge-flip and non-prompt leptons.

\subsection{Background from $WZ$ production}
\label{sec:WZ}
\begin{figure}[t!]
\subfloat[]{
\includegraphics[width=0.48\textwidth]{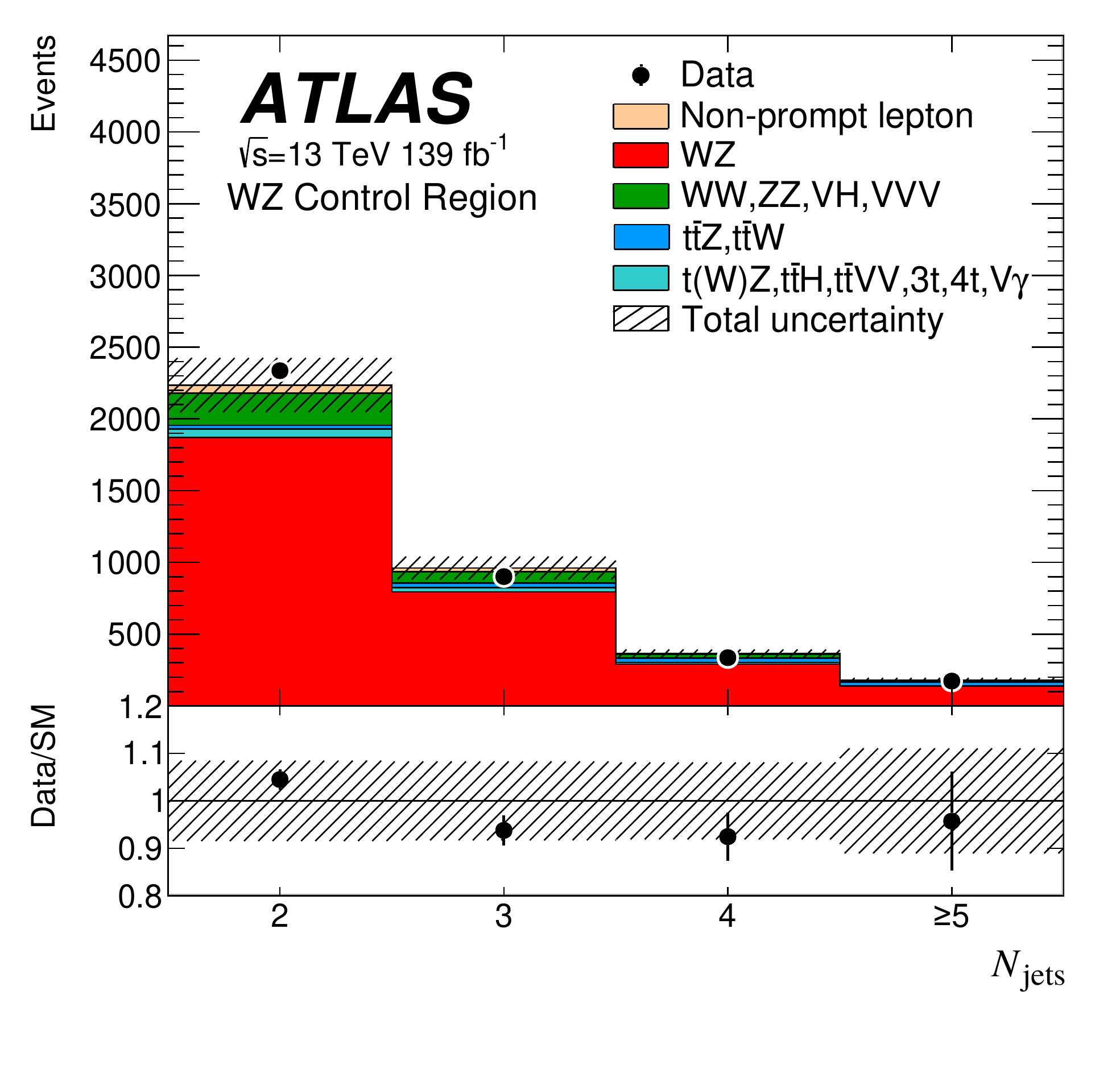}
 
\label{fig,WZ CR}
}
\subfloat[]{
\includegraphics[width=0.48\textwidth]{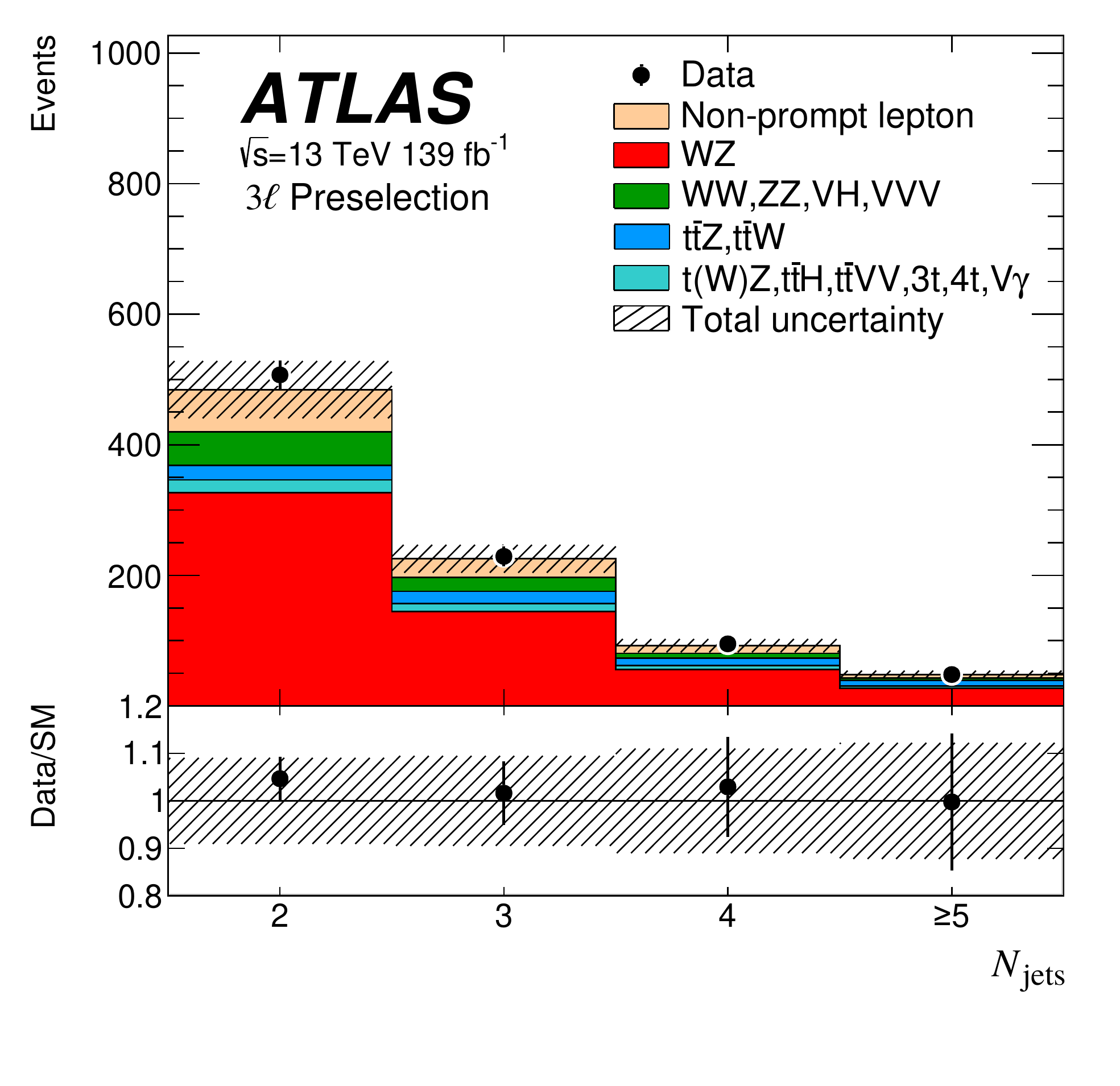}
 
\label{fig,Njets 3L}
}
\caption{
Distribution of the jet multiplicity in the (a) $WZ$ control region and in the (b) \lll\ preselection region.
The data (dots) are compared with the estimated contributions from the background sources (histograms).
The normalisation factor is applied to the $WZ$ MC contribution. The last bin includes overflows.
In each figure the bottom panel shows the ratio of data to the estimated background. The hatched band around unity represents the total uncertainty of the background.
}
\end{figure}

The $WZ$ process is a dominant source of background in the \ll and \lll SRs.
To correct a mismodelling seen in the jet multiplicity distribution~\cite{Aaboud:2019gxl}, a normalisation factor
is computed and then applied to the $WZ$ background events containing two or more jets.
The normalisation factor is measured in a dedicated $WZ$ control region.
It is selected by requiring exactly three tight leptons with $\pt > \SI{20}{GeV}$, at least two jets and no $b$-jet in the event.
Finally, there must be at least one pair of same-flavour opposite-charge leptons with a mass compatible with the $Z$-boson mass ($|m_{\ell\ell}^\text{oc}- m_{Z}| < 10$~\GeV).
The latter criterion ensures that the $WZ$ control region and the \lll SRs are orthogonal.

The normalisation factor is derived in a fit of a first-order polynomial, as a function of jet multiplicity, to ratios of the
data event yield (subtracting all non-$WZ$ contributions) to the $WZ$ event yield predicted in MC simulation.
The value of the polynomial at $N_{\mathrm{jets}} =0$ is used to scale the predicted $WZ$ yields and found is to be
$0.83\pm0.07$, where the uncertainty includes the statistical and systematic components.
The different sources of systematic uncertainty are discussed in Section~\ref{sec:syst_unc}.
The jet multiplicity distribution in the $WZ$ control region is shown in Figure~\ref{fig,WZ CR}, and
for illustration the distribution in the \lll\ preselection region is also shown in Figure~\ref{fig,Njets 3L}.
The normalisation factor is applied to the $WZ$ MC contribution.
The sum of estimated backgrounds agrees with data within the assigned systematic uncertainties.

\subsection{Electron charge-flip background}
In the \ll channel, a background contribution is expected from events with opposite-charge lepton pairs where the charge of one of the leptons is misidentified.
The charge-flip background is only significant for electrons and is mainly due to interactions of the electron with the ID material.
 
The misidentification rate is measured using a large data sample of
dielectron events originating mainly from $Z\rightarrow e^+e^-$ decays selected by requiring two tight electrons with an invariant mass between 80 and \SI{100}{GeV}.
The sample contains mostly opposite-charge dielectron pairs, with a small fraction of same-charge dielectron pairs.
The fraction of same-charge dielectron events is used to extract the charge misidentification rate
as a function of the electron \pt and $\eta$, using the method described in Ref.~\cite{EGAM-2018-01}.
This rate is found to range between $0.01$\% and $4$\%, where higher values
are obtained at large rapidities due to the larger amount of material traversed by the electrons.
The statistical uncertainty of this estimate varies between 2\% and 26\% and is taken as a systematic uncertainty in the charge misidentification rate.
The background in both the opposite-charge and same-charge samples
is estimated in a cubic polynomial function fit of the high (100--\SI{120}{GeV}) and low (60--\SI{80}{GeV}) $m_{\ell\ell}$ sidebands.
The uncertainty in the background is estimated by varying the sidebands and propagates to a 3\% uncertainty in the estimated charge misidentification rate.
The final systematic uncertainty combines all the sources mentioned.
 
The charge-flip background in a given region is estimated by selecting data events with opposite-charge dilepton pairs, but otherwise identical selection,
and weighting them by the probability that the charge of the electrons is misidentified.
Another source of systematic uncertainty is estimated by comparing, in simulated $V$+jets, $t\bar{t}$, and $WW$ events,
the number of same-charge events estimated from opposite-charge events with the prediction; it accounts for differences between the charge misidentification rates in different processes and is found to be approximately 10\%.
For this test, the misidentification rate was measured using $Z\rightarrow e^+e^-$ MC simulations, using the same method as for the measurement performed with data.
 
To estimate the electron charge-flip background in the regions defined with loose electrons the methodology described above is applied.
For this estimation, dedicated electrons misidentification rates measured with a $Z\rightarrow e^+e^-$ sample selected by requiring one tight and one loose electrons are used.

\subsection{Non-prompt-lepton background}
\label{sec:fakeLeptons}
The composition of the non-prompt-lepton background in the SRs varies considerably among the analysis channels.
Therefore, the methods to estimate these contributions are different for the \ll, \lll and \llll channels.
In the \ll and \lll channels the non-prompt-lepton background is estimated using a fake-factor method~\cite{Aaboud:2018qcu},
while the simulation is corrected with scale factors measured in data for the \llll channel.

\paragraph{Non-prompt-lepton background estimate for the \ll channel}
The estimate of the background from non-prompt leptons assumes that these contributions can be extrapolated from a control region enriched in non-prompt leptons with a so-called fake factor.
The control region is selected using the kinematic requirements of the preselection or the signal regions but alternative lepton selection criteria.
The latter means that at least one of the selected leptons is required to satisfy the loose but not the tight lepton requirements.
 
The fake factors are calculated separately for electrons and muons in control regions with kinematic selections designed to enhance their content in non-prompt leptons.
This is achieved by applying the preselection requirements of the \ll channel, except that $\MET$ must be lower than \SI{70}{GeV}.
Only events with electron and muon same-charge pairs are then used.
The fake factor is defined as ratio of the number of events in the control region with all the selected leptons required to pass the tight signal requirements,
to the number of events in the same region but with one of the selected leptons satisfying alternative lepton requirements.
The measurement is performed as a function of the lepton \pt.
The fake factor dependency on the lepton $\eta$ was also checked and found to be negligible.
The SM processes with prompt leptons and the charge-flip contributions are subtracted in the control region.
For the electron and muon fake-factor measurements, the lepton with the second highest $\pt$ is assumed to be the non-prompt one.
 
The measured fake factors are $\numRP{0.028}{2}\pm \numRP{0.006}{2}$
for electrons and muons up to $\pt = \SI{40}{GeV}$,
and increase to $\numRP{0.160}{2}\pm \numRP{0.051}{2}$ and $\numRP{0.085}{2}\pm \numRP{0.024}{2}$, respectively, for electrons and muons with $\pt>\SI{60}{GeV}$.
The uncertainties are statistical only.
A systematic uncertainty of 20\% (10\%) in the electron (muon) fake factor
is estimated by studying the variation of the fake factor with \met.
For this measurement two \met bins are considered, $< \SI{70}{GeV}$ and $> \SI{70}{GeV}$.
This uncertainty accounts for the different compositions of the non-prompt leptons in the control region and the SRs.
Another source of systematic uncertainty accounts for how often the non-prompt lepton is actually the one with the highest lepton $\pt$ and not
the one with the second highest $\pt$, as assumed.
It is estimated from generator-level studies performed with MC simulations.
This source is dominant in the region with the lepton $\pt > \SI{60}{GeV}$,
where the uncertainty reaches 45\% (80\%) for electrons (muons).
Uncertainties in the SM processes with prompt leptons (approximately $20\%$) and electron charge-flip ($15\%$) background subtraction are also included.
The overall uncertainty amounts to approximately 30\% (20\%) and
55\% (85\%) for the fake factors for electrons (muons) with  $20<\pt<\SI{60}{GeV}$ and $\pt>\SI{60}{GeV}$, respectively.

\paragraph{Non-prompt-lepton background estimate for the \lll channel}
The same method as the one employed for the \ll channel is used in the \lll channel.
Here the opposite-charge lepton $\ell_0$ passes the loose selection, and it is assumed to be prompt,
an assumption that is found to be valid in MC simulation.
The control region used to calculate the fake factors uses the \lll preselection requirements,
except that exactly one jet is required.
Only events with electron and muon same-charge pairs are then used.
The muon fake factor is found to be $\numRP{0.032}{2}\pm \numRP{0.009}{2}$ and the electron fake factor is found to be $\numRP{0.021}{2} \pm \numRP{0.009}{2}$,
where the uncertainties are statistical only.
A systematic uncertainty of 15\% is estimated by measuring the lepton fake factor in a control region enriched in events from \ttbar production.
This uncertainty accounts for the different compositions of the non-prompt leptons in the control region and the SRs.
Uncertainties in the subtraction of the SM processes with prompt leptons are found to be approximately 55\% (45\%) for the electron (muon) fake factor.
Another 20\% uncertainty comes from generator-level studies performed with MC simulations to test the assumption that $\ell_0$ is a prompt lepton.
The fake factors' dependency on the lepton \pt is also studied, and the deviations from the nominal fake factors are within the statistical uncertainty.
When all sources of systematic uncertainty are combined,
the total systematic uncertainty of the electron (muon) fake factor is found to be 60\% (50\%).

\paragraph{Non-prompt-lepton background estimate for the \llll channel}
\begin{table}[t!]
\begin{center}
\def\arraystretch{1.2}
\caption{
The selection criteria that define the control regions enriched in non-prompt leptons used to determine the MC scale factors for the \llll channel.
The leptons are required to pass the loose and minimally isolated (L$^{*}$) requirements.
The transverse mass, $m_\text{T}$, is calculated as the invariant mass of the vector sum of the transverse momentum of the non-prompt-lepton candidate and the missing transverse momentum.
The symbol ``--'' means no requirement is made.
The equal sign $(=)$ is used to emphasise that the selection criterion has to be exactly the given number.
}
\label{Tab:CR_Fakes}
 
\scriptsize
\resizebox{0.66\textwidth}{!}
{
\begin{tabular}{l||c|c}
\hline\hline
Sample          & $Z$+jets-enriched region & \ttbar-enriched region \\
\hline\hline
 
$N_\ell$ (type L$^{*}$)       & =3                         & =3        \\
$|\sum Q_\ell|$         & 1                          & 1 		   \\
\nj                     & $1 \; \mathrm{or} \;2 $    & $1 \; \mathrm{or} \; 2 $ \\
$\pt^{\mathrm{jet}}$    & $>25$~\GeV                 & $>30$ (25)~\GeV\\
$Z$-window              &$|m_{\ell\ell}^\text{oc}- m_{Z}| < 10$~\GeV & No same-flavour opposite-charge $\ell$ pair\\
$\MET$                  & $<50$~\GeV                 & --\\
$m_\text{T}$            & $<50$~\GeV                 & -- \\

\hline\hline
\end{tabular}
}
\end{center}
\end{table}

There are too few data events to use the fake-factor method in the \llll channel.
Instead, the non-prompt-lepton yields are estimated from simulation corrected
using scale factors that are extracted in control regions enriched in non-prompt leptons.
The main sources of non-prompt leptons are $b$- and $c$-hadron decays.
For electrons a small component from light-flavour hadrons is also present, while for muons this component is negligible.
The scale factor is measured separately for non-prompt leptons from $b$-hadron decays and light-flavour hadrons.
 
Two data samples enriched in non-prompt leptons originating from \zj and \ttbar events
are used to study how well the MC simulations describe the non-prompt leptons originating from light-flavour (LF) and heavy-flavour (HF) hadrons, respectively.
These control regions are defined in Table~\ref{Tab:CR_Fakes}.
Three scale factors, $\lambda_\mathrm{HF}^{e}$, $\lambda_\mathrm{LF}^{e}$ and $\lambda_\mathrm{HF}^{\mu}$,
are obtained by solving the system of equations:
\begin{align}
N^{e}_{\mathrm{Data} | Z\text{+jets}}-N^{e}_{\mathrm{Prompt} | Z\text{+jets}}  &= \lambda_\mathrm{HF}^{e} N^{e}_{\mathrm{HF} | Z\text{+jets}} + \lambda_\mathrm{LF}^{e} N^{e}_{\mathrm{LF} | Z\text{+jets}}  \; \; ,  \nonumber \\
N^{e}_{\mathrm{Data} | t\overline{t}}-N^{e}_{\mathrm{Prompt} | t\overline{t}}  &= \lambda_\mathrm{HF}^{e} N^{e}_{\mathrm{HF} | t\overline{t}} + \lambda_\mathrm{LF}^{e} N^{e}_{\mathrm{LF} | t\overline{t}}  \; \; ,  \nonumber \\
N^{\mu}_{\mathrm{Data} | t\overline{t}} - N^{\mu}_{\mathrm{Prompt} | t\overline{t}} - N^{\mu}_{\mathrm{LF} | t\overline{t}}  &= \lambda_\mathrm{HF}^{\mu} N^{\mu}_{\mathrm{HF} | t\overline{t}} \; \; , \nonumber
\end{align}
with the event yields $N^\ell$, $\ell=e,\mu$ in data (Data),
and the background from prompt leptons (Prompt), light-flavour leptons (LF) and heavy-flavour leptons (HF),
in the $t\overline{t}$-enriched ($t\overline{t}$) and the $Z$+jets-enriched ($Z$+jets) control regions.
 
The scale factors are measured to be
$\lambda_\mathrm{HF}^{e} = 0.98 \pm 0.18 $,
$\lambda_\mathrm{LF}^{e} = 1.34 \pm 0.17 $ and
$\lambda_\mathrm{HF}^{\mu} = 0.94 \pm 0.04 $, where the uncertainties are statistical only.
Systematic uncertainties of 5\%, 4\% and 2\% for the $\lambda_\mathrm{HF}^{e}$, $\lambda_\mathrm{LF}^{e}$ and $\lambda_\mathrm{HF}^{\mu}$ scale factors, respectively, are estimated
by varying the jet multiplicity and the lepton \pt threshold in the nominal $Z$+jets- and \ttbar-enriched control regions.
These uncertainties account for the different compositions of the non-prompt leptons in the control regions and the SRs.
Uncertainties in the prompt-lepton subtraction are dominated by theoretical uncertainties.
The total systematic uncertainties combine all the sources mentioned and
are approximately 6\%, 15\% and 4\% for the $\lambda_\mathrm{HF}^{e}$, $\lambda_\mathrm{LF}^{e}$ and $\lambda_\mathrm{HF}^{\mu}$ scale factors, respectively.

\subsection{Validation}
 
\begin{figure}[t!]
\centering
\subfloat[]{
\includegraphics[width=0.49\textwidth]{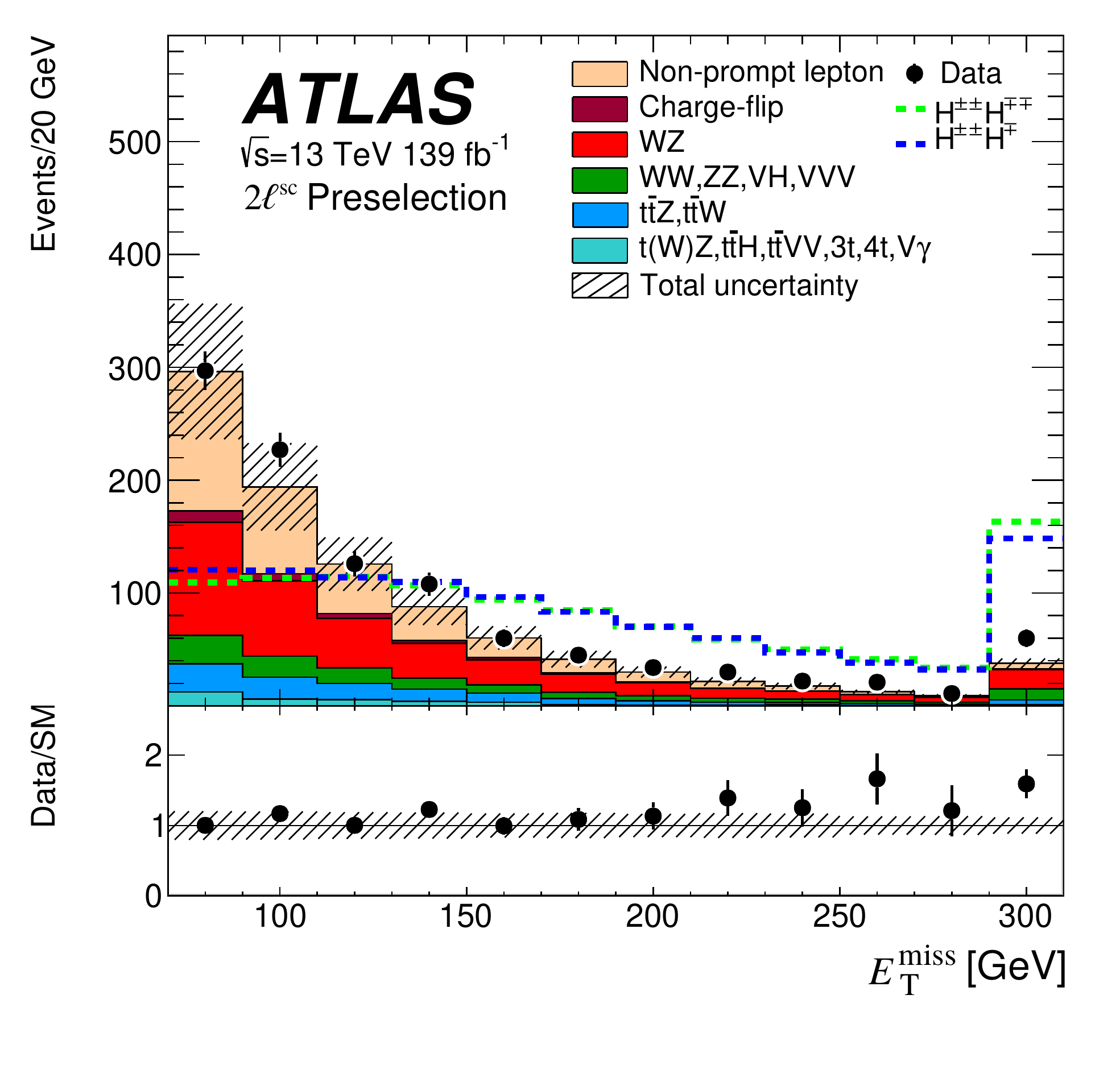}
}
\subfloat[]{
\includegraphics[width=0.49\textwidth]{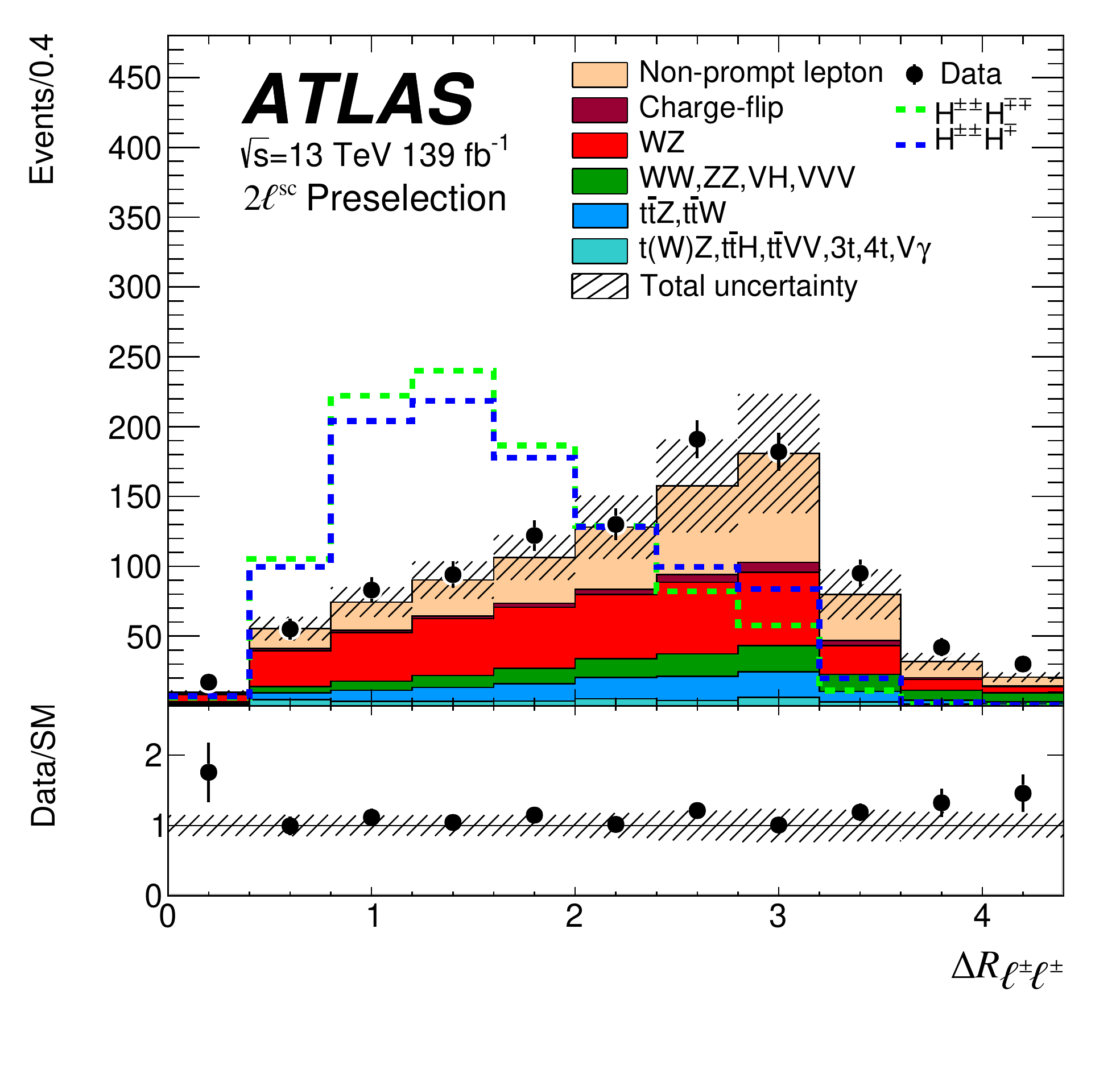}
}
\hspace{0.cm}
\subfloat[]{
\includegraphics[width=0.49\textwidth]{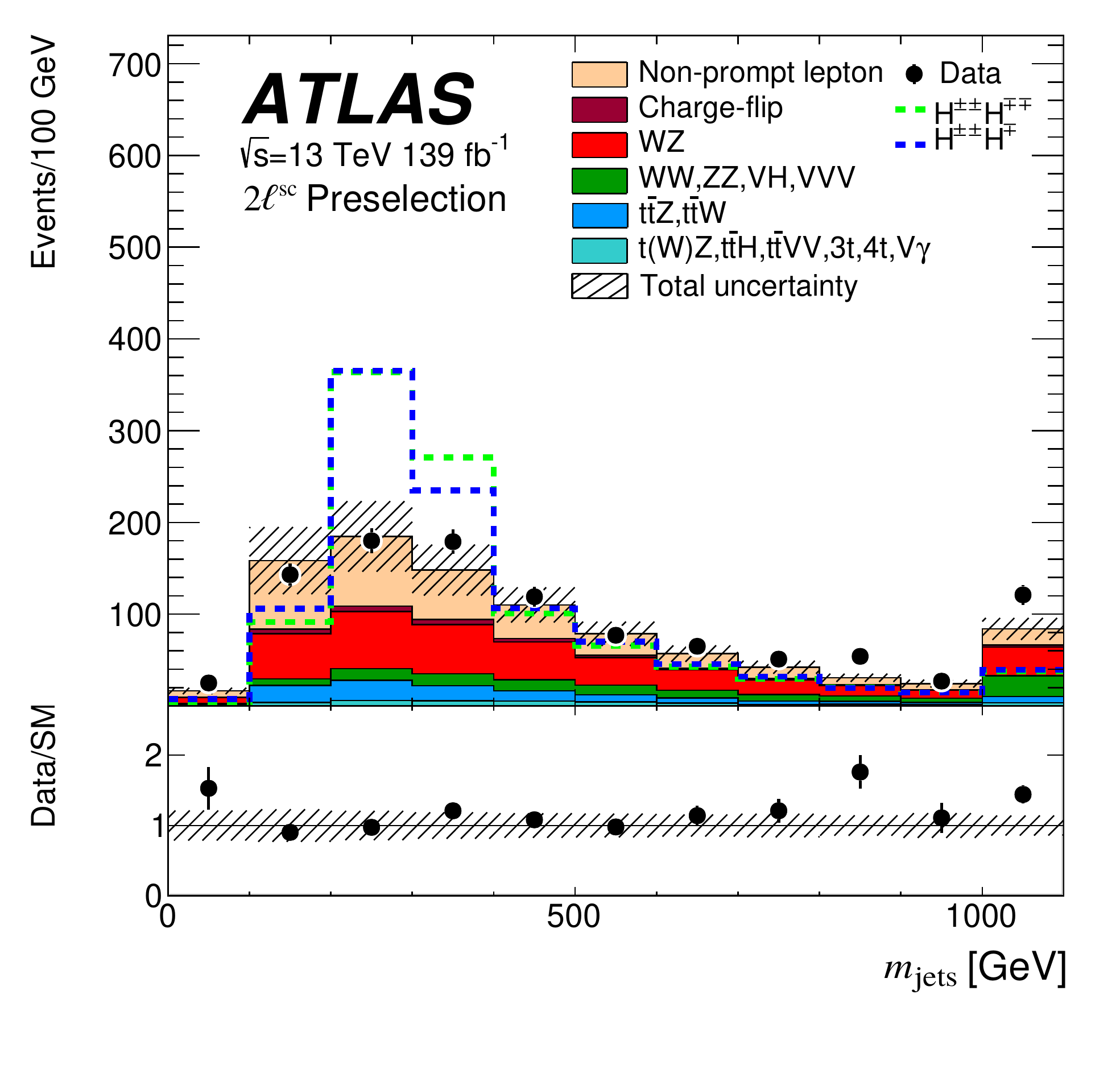}
}
\subfloat[]{
\includegraphics[width=0.49\textwidth]{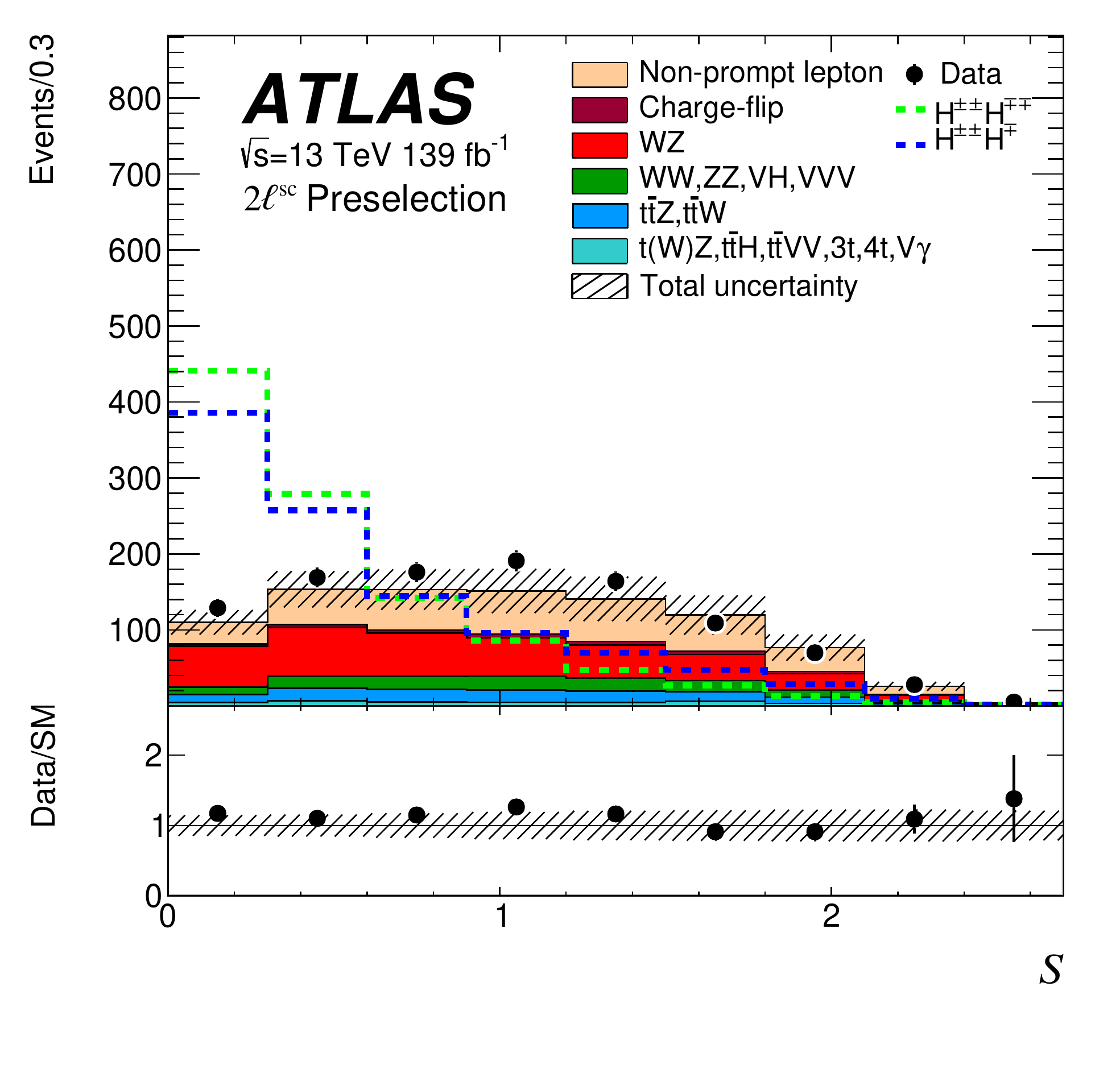}
}
\caption{
Distribution of selected variables used to define the \ll SRs.
The events are selected with the preselection requirements listed in Table~\ref{Tab:preselection}.
The data (dots) are compared with the expected contributions from the relevant background sources (histograms).
The expected signal distributions for $m_{H^{\pm\pm}} = \SI{300}{GeV}$ are also shown, scaled to the observed number of events.
The last bin includes overflows.
In each figure the bottom panel shows the ratio of data to the estimated background. The hatched band around unity represents the total uncertainty of the background.
}
\label{fig,2L_PRES}
\end{figure}
 
\begin{figure}[t!]
\centering
\subfloat[]{
\includegraphics[width=0.49\textwidth]{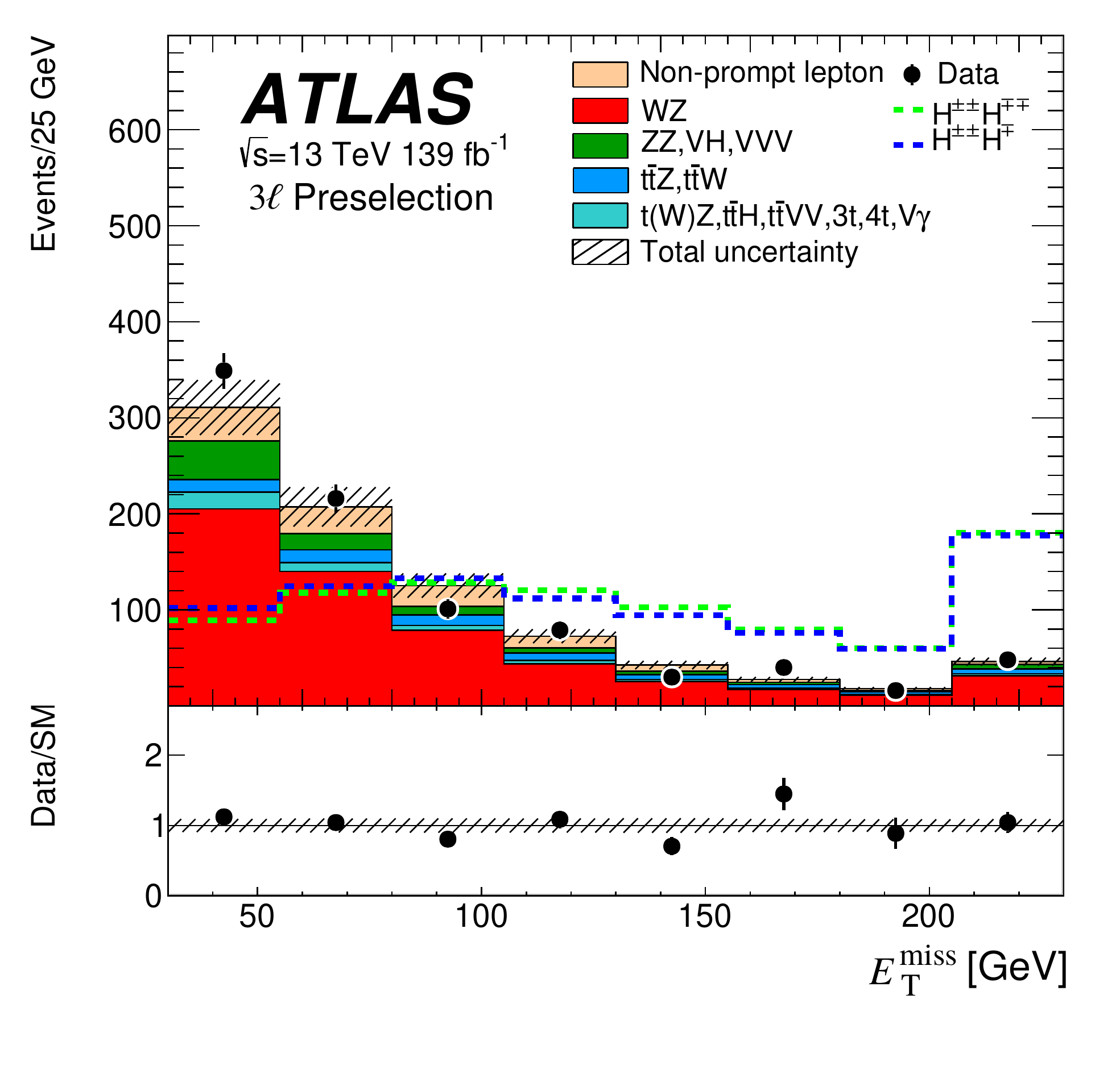}
}
\subfloat[]{
\includegraphics[width=0.49\textwidth]{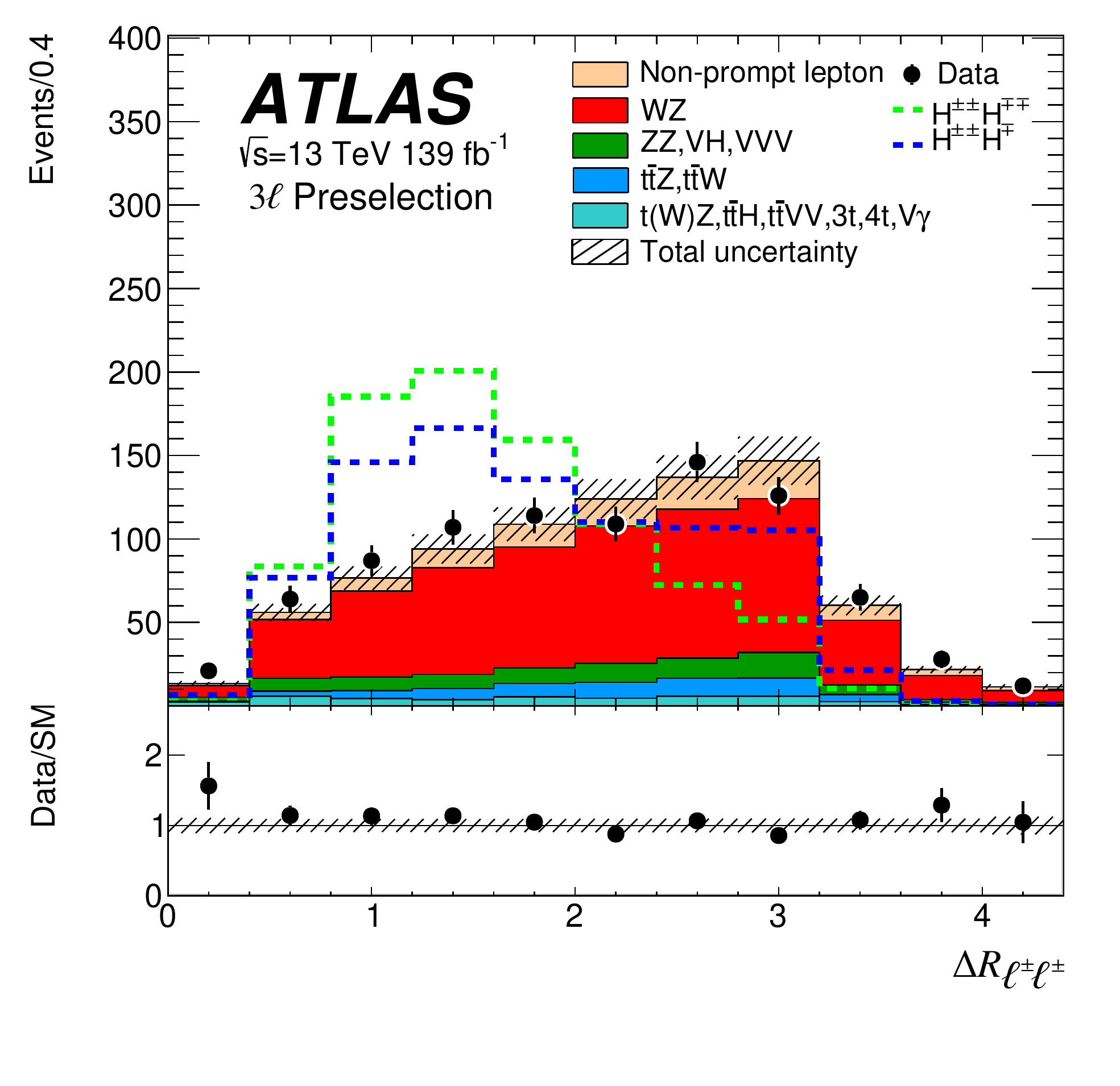}
}
\hspace{0.cm}
\subfloat[]{
\includegraphics[width=0.49\textwidth]{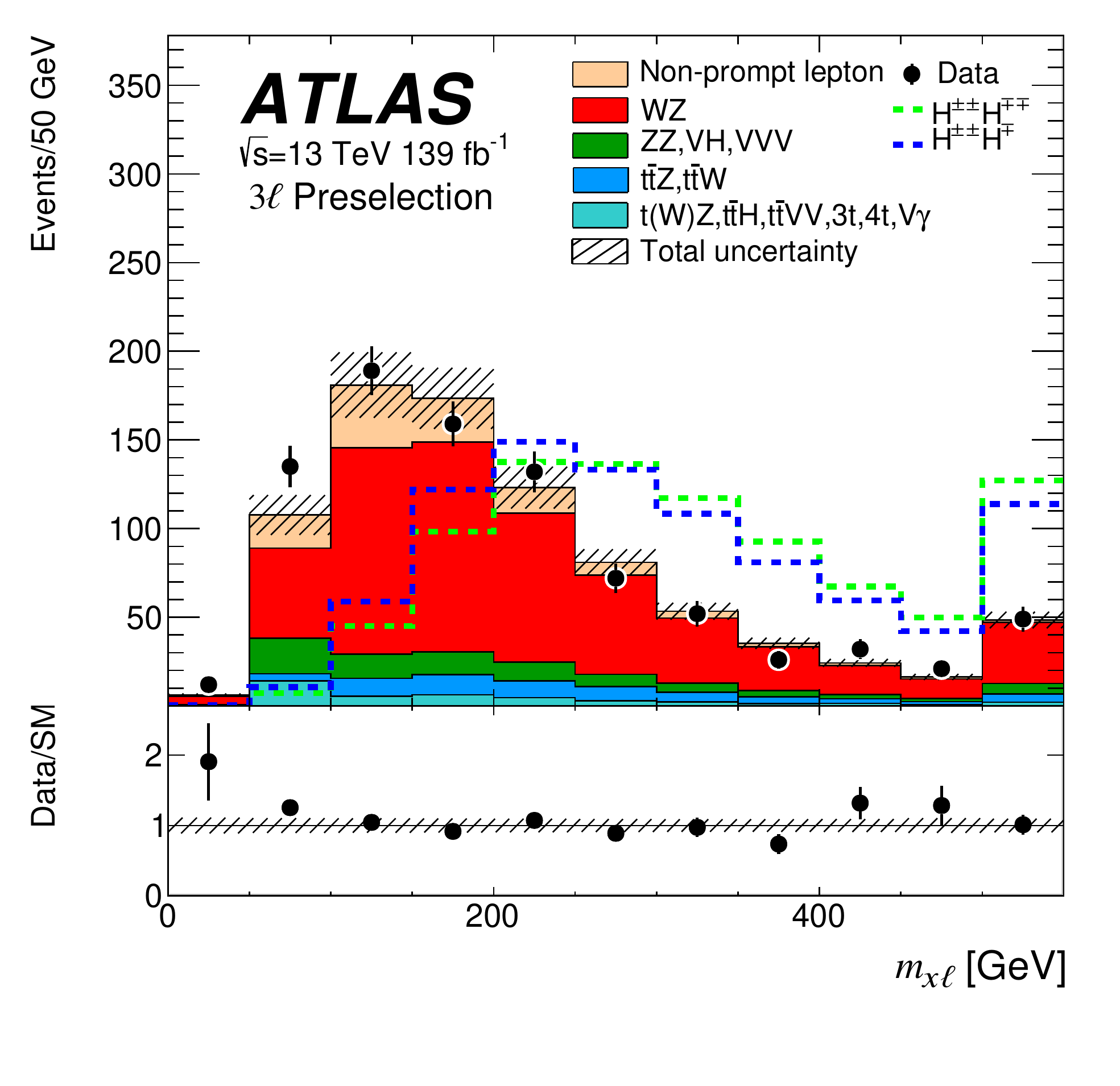}
}
\subfloat[]{
\includegraphics[width=0.49\textwidth]{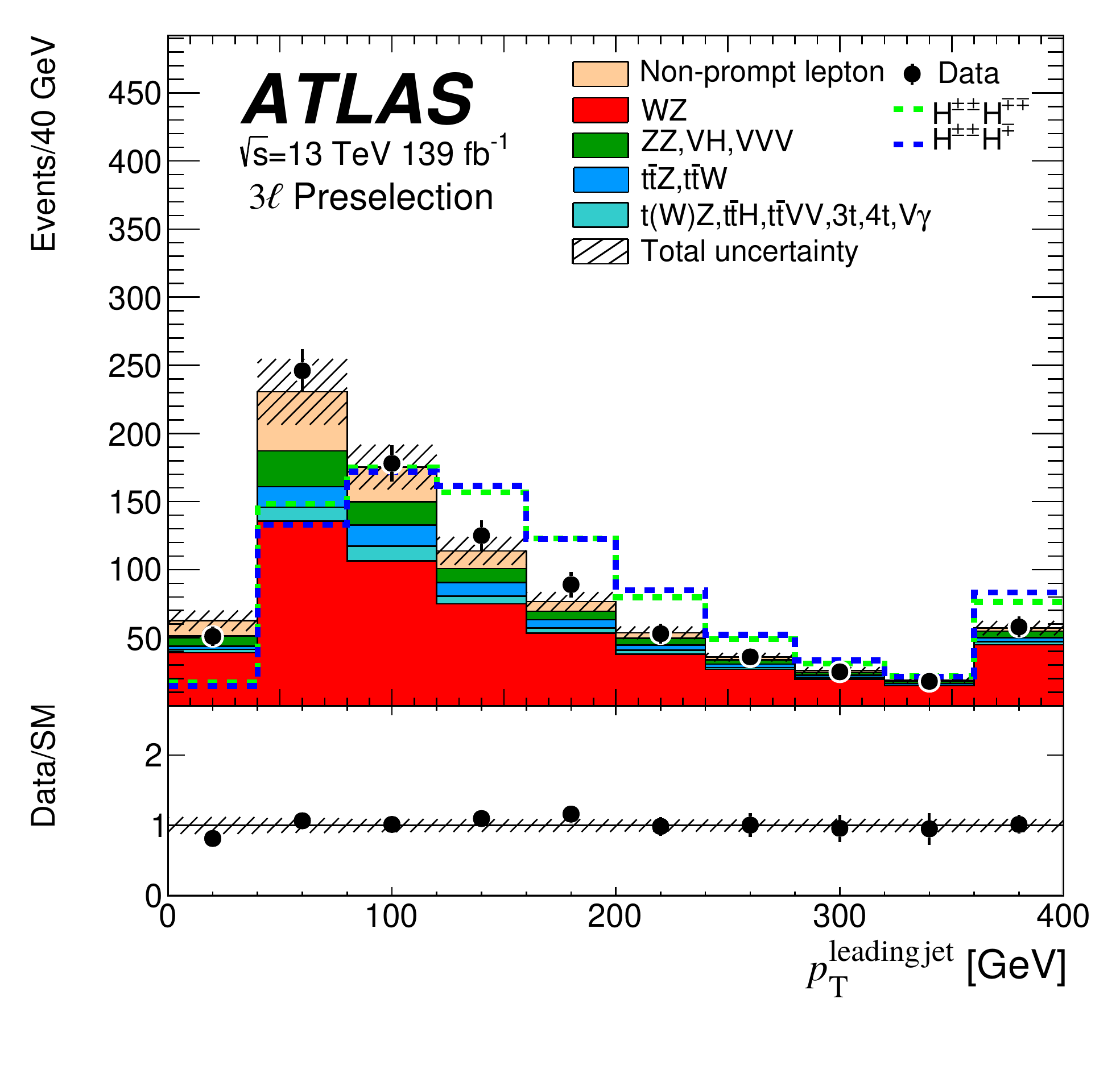}
}
\caption{
Distribution of selected variables used to define the \lll SRs.
The events are selected with the preselection requirements listed in Table~\ref{Tab:preselection}.
The data (dots) are compared with the expected contributions from the relevant background sources (histograms).
The expected signal distributions for $m_{H^{\pm\pm}} = \SI{300}{GeV}$ are also shown, scaled to the observed number of events.
The last bin includes overflows.
In each figure the bottom panel shows the ratio of data to the estimated background. The hatched band around unity represents the total uncertainty of the background.
}
\label{fig,3L_PRES}
\end{figure}
 
\begin{figure}[t!]
\centering
\subfloat[]{
\includegraphics[width=0.49\textwidth]{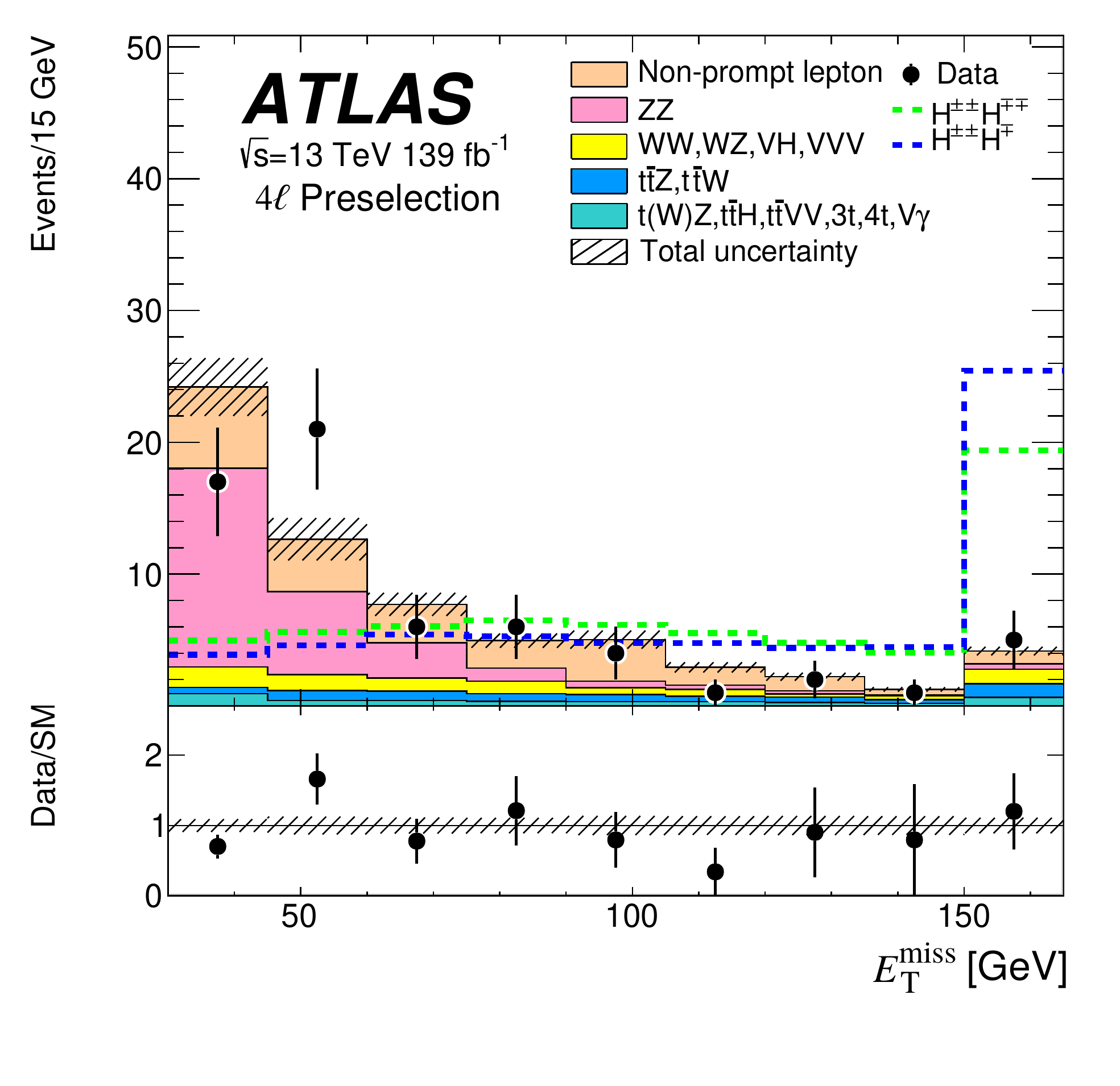}
}
\subfloat[]{
\includegraphics[width=0.49\textwidth]{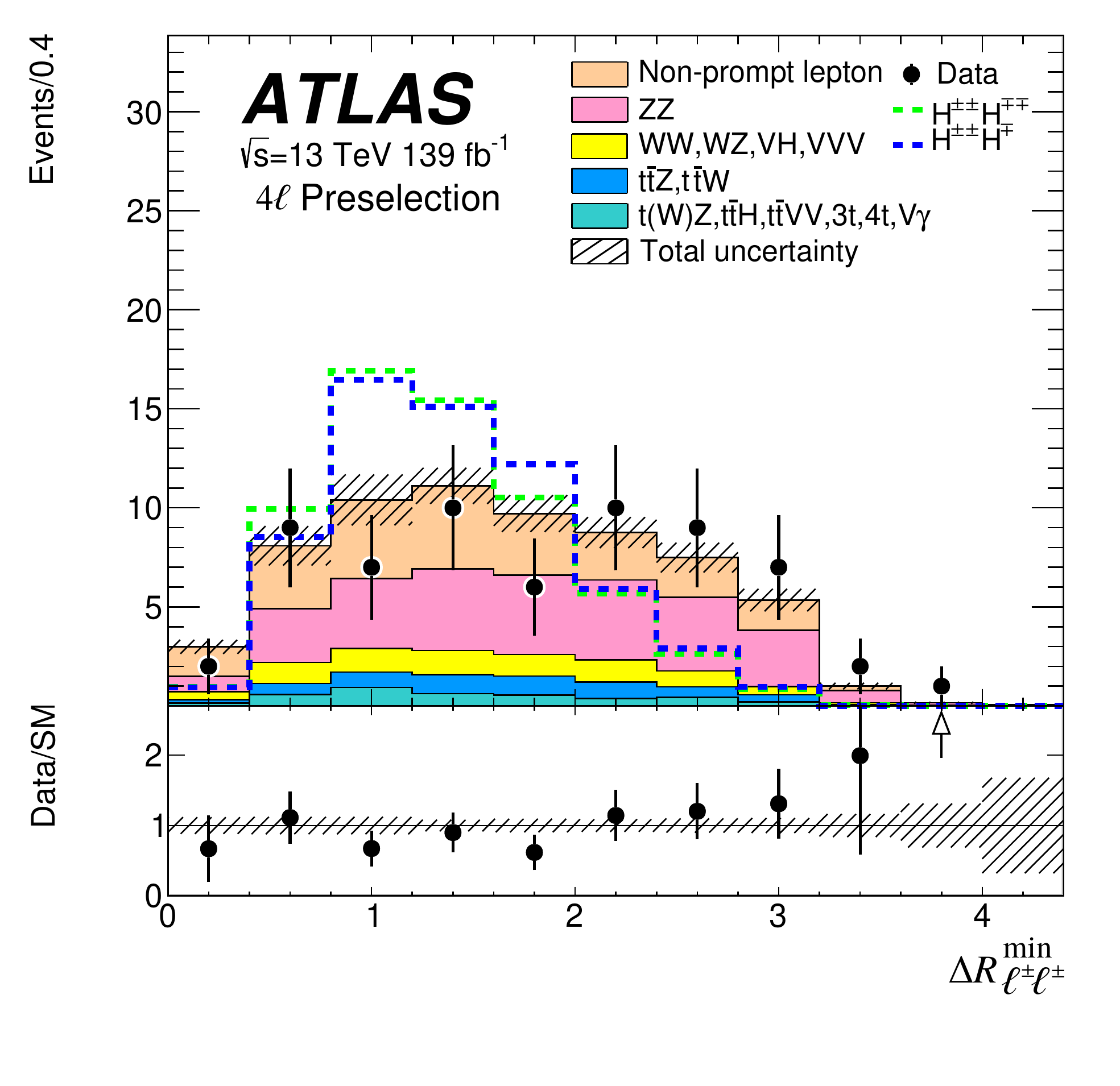}
}
\hspace{0.cm}
\subfloat[]{
\includegraphics[width=0.49\textwidth]{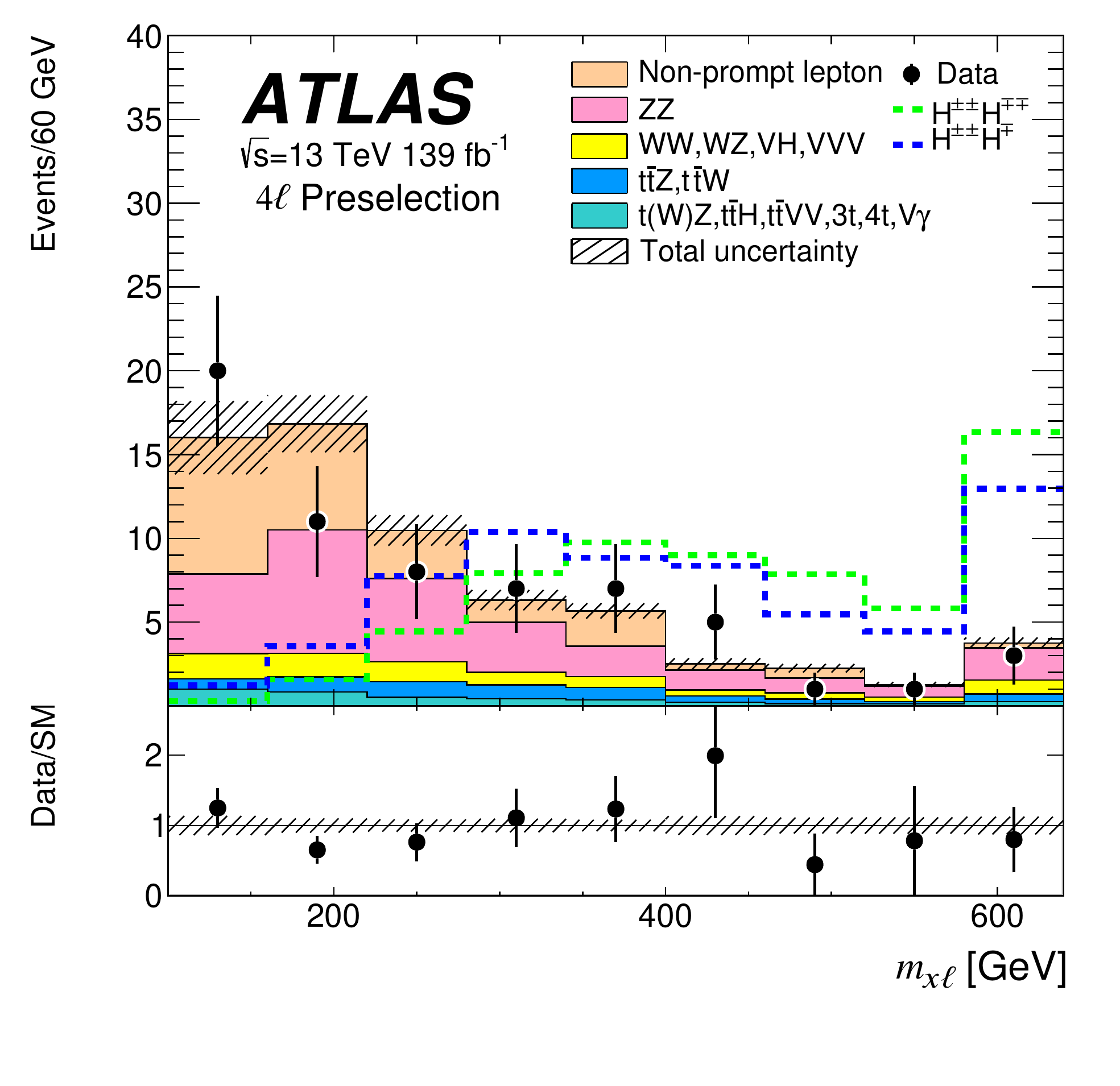}
}
\subfloat[]{
\includegraphics[width=0.49\textwidth]{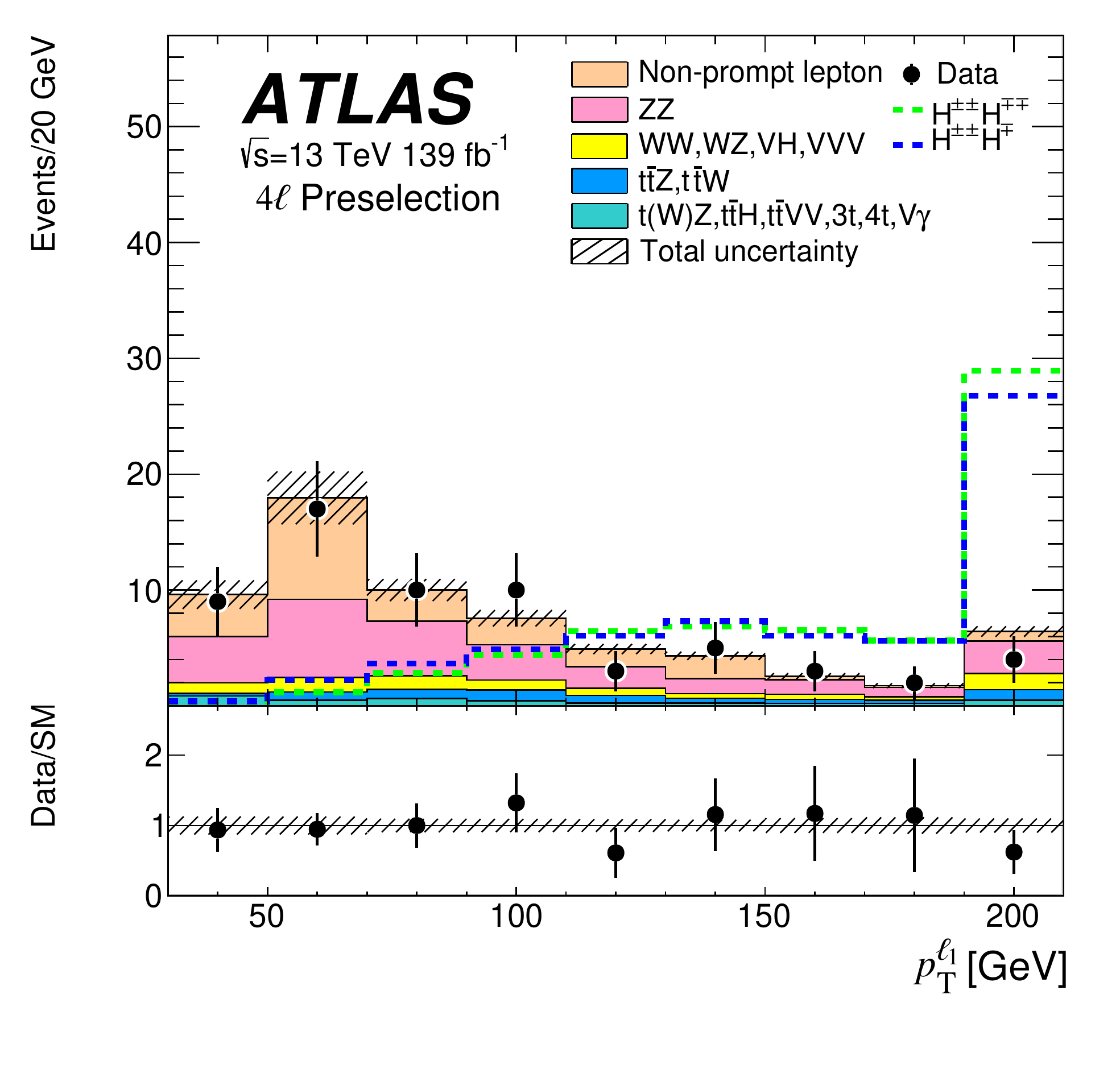}
}
\caption{
Distribution of selected variables used to define the \llll SRs.
The events are selected with the preselection requirements listed in Table~\ref{Tab:preselection}.
The data (dots) are compared with the expected contributions from the relevant background sources (histograms).
The expected signal distributions for $m_{H^{\pm\pm}} = \SI{300}{GeV}$ are also shown, scaled to the observed number of events.
The last bin includes overflows.
In each figure the bottom panel shows the ratio of data to the estimated background. The hatched band around unity represents the total uncertainty of the background.
}
\label{fig,4L_PRES}
\end{figure}

The data-based methods employed to estimate the backgrounds are validated by comparing
the event yields in data with the combined predictions for these backgrounds, added to
MC predictions for SM processes with prompt signal leptons.
Distributions of selected variables in the \ll, \lll and \llll channels are shown
in Figures~\ref{fig,2L_PRES},~\ref{fig,3L_PRES} and~\ref{fig,4L_PRES}, respectively,
after the preselection requirements from Table~\ref{Tab:preselection} are applied.
Good agreement is observed in both normalisation and shape, demonstrating that the background contributions are well estimated.
The expected distributions of both signal models are shown for $m_{H^{\pm\pm}} = \SI{300}{GeV}$ to illustrate the discrimination power of the selected variables.

The signal contamination at the preselection level was studied for $m_{H^{\pm\pm}} = \SI{300}{GeV}$ and $m_{H^{\pm\pm}} = \SI{200}{GeV}$ mass hypotheses,
corresponding to the $H^{\pm\pm}$ pair production mode and the $H^{\pm\pm}$ and $H^{\pm}$ associated production mode, respectively.
These mass hypotheses were selected because they are close to the sensitive mass range of the current analysis.
A maximum value of 8.5\% (2\%) $H^{\pm\pm} H^{\mp\mp}$ ($H^{\pm\pm} H^{\mp}$) signal contamination was found in all the individual \ll, \lll and \llll channels at preselection level.

\section{Systematic uncertainties} 
\label{sec:syst_unc}
 
Uncertainties in the signal and background yields arise from experimental uncertainties and from the theoretical accuracy of the prediction of the SM background yields.
The experimental uncertainties arise from
the luminosity determination, modelling of pile-up interactions,
the reconstruction and identification of electrons, muons and jets and from the uncertainties associated with the
data-based methods that are used to estimate the non-prompt lepton and charge-flip electron backgrounds.
 
The uncertainty in the combined 2015--2018 integrated luminosity is 1.7\%~\cite{ATLAS-CONF-2019-021}, obtained using the LUCID-2 detector~\cite{LUCID2} for the primary luminosity measurements.
Uncertainties in the modelling of pile-up interactions are estimated by reweighting the distribution of the
average number of $pp$ interactions per bunch crossing in the simulation, such that the average number of interactions changes by $\pm9\%$.
The impact of these uncertainties on the background event yields estimated from MC simulation is found to be lower than 2\%.
 
The uncertainties related to event reconstruction include the lepton~\cite{PERF-2015-10,PERF-2017-03}
and jet \cite{JETM-2018-05} energy scales and resolutions,
and the uncertainties of the \MET soft term~\cite{PERF-2016-07}.
The overall impact on the background from SM processes with prompt leptons and the signal yields in the SRs from these systematic uncertainties is found to be lower than {10\%}.
The dominant contribution comes from the jet energy scale component.
 
The uncertainties in the efficiencies of the electron \cite{PERF-2017-01} and muon \cite{PERF-2015-10} reconstruction, identification and trigger are also included.
Their impact on the estimated yields in the SRs for the signal and the background from SM processes with prompt leptons is lower than {$5$\%}. The uncertainties in the $b$-jet identification are found to be negligible.
 
The uncertainties in the background yield from electron charge-flip are
in the range of $20$--$30$\% for the \ll $ee$ and $e\mu$ channels.
The uncertainties in the non-prompt-lepton contributions in the SRs are between 25\% and 90\%
and include the uncertainties described in Section~\ref{sec:fakeLeptons}, and statistical uncertainties in the control regions where the contributions are estimated.

For the background from $ZZ$, $WW$ and $VVV$ multiboson processes, and $V\gamma$, \tZ and \ttX\ processes (\ttW, \ttZ, \ttH), the systematic uncertainties
due to higher-order QCD corrections are evaluated by varying the renormalisation and factorisation scales
independently by factors of two and one-half, and removing combinations where the variations differ by a factor of four.
The uncertainties due to the PDF and the $\alpha_\mathrm{s}$ value used in the PDF determination are evaluated using the {PDF4LHC} prescription~\cite{Butterworth:2015oua}.
In the SRs defined for $m_{H^{\pm\pm}} = \SI{300}{GeV}$, the theory uncertainty in the background yields from
\ZZ and $WW$ processes varies between {15\%} and {40\%}.
For the background from $VVV$ it is approximately {10\%},
for the background from $V\gamma$ it is approximately 35\%
and for the \tZ\ and \ttX\ processes it is approximately 14\%.
 
The total uncertainty in the estimated $WZ$ yields is {9}\%
and includes the statistical uncertainty as well as two sources of theoretical uncertainty.
Systematic uncertainties due to higher-order QCD corrections are evaluated using the same prescription as for the other diboson processes, and are found to be 3\%.
The second source is evaluated by comparing the results obtained with the linear fit model (Section~\ref{sec:WZ}) in the $WZ$ control region,
and in a region defined with the $WZ$ control region requirements except that three jets must be in the event.
This uncertainty is found to be 8.4\%.
To validate the assigned uncertainty, several checks are performed.
The linear fit function is changed to a quadratic one to check the quality of the fit model.
The fit parameterisation is studied in the $WZ$-enriched region
and in a region defined with the $WZ$ control region requirements except that three jets must be in the event.
The differences between the obtained results are found to be covered by the total uncertainty.
The choice of parton shower model is studied in the $WZ$ control region by comparing event samples simulated with \sherpa~v2.2.1 and \mgamc,
and the differences are found to be covered by the total uncertainty.
For the $WZ$ background in regions with lower jet multiplicities, the uncertainty is estimated with the same methodology as for the other diboson processes described above.

An uncertainty of 50\% is assigned to the other backgrounds (\ttt, \tttt, \ttWW and $VH$),
This large value is assigned to cover uncertainties from missing higher-order corrections and the PDF sets.
Since these processes produce a larger number of jets at the first order of the perturbative expansion, they are less sensitive to parton shower modelling uncertainties.

\begin{figure}[t!]
\centering
\includegraphics[width=0.8\textwidth]{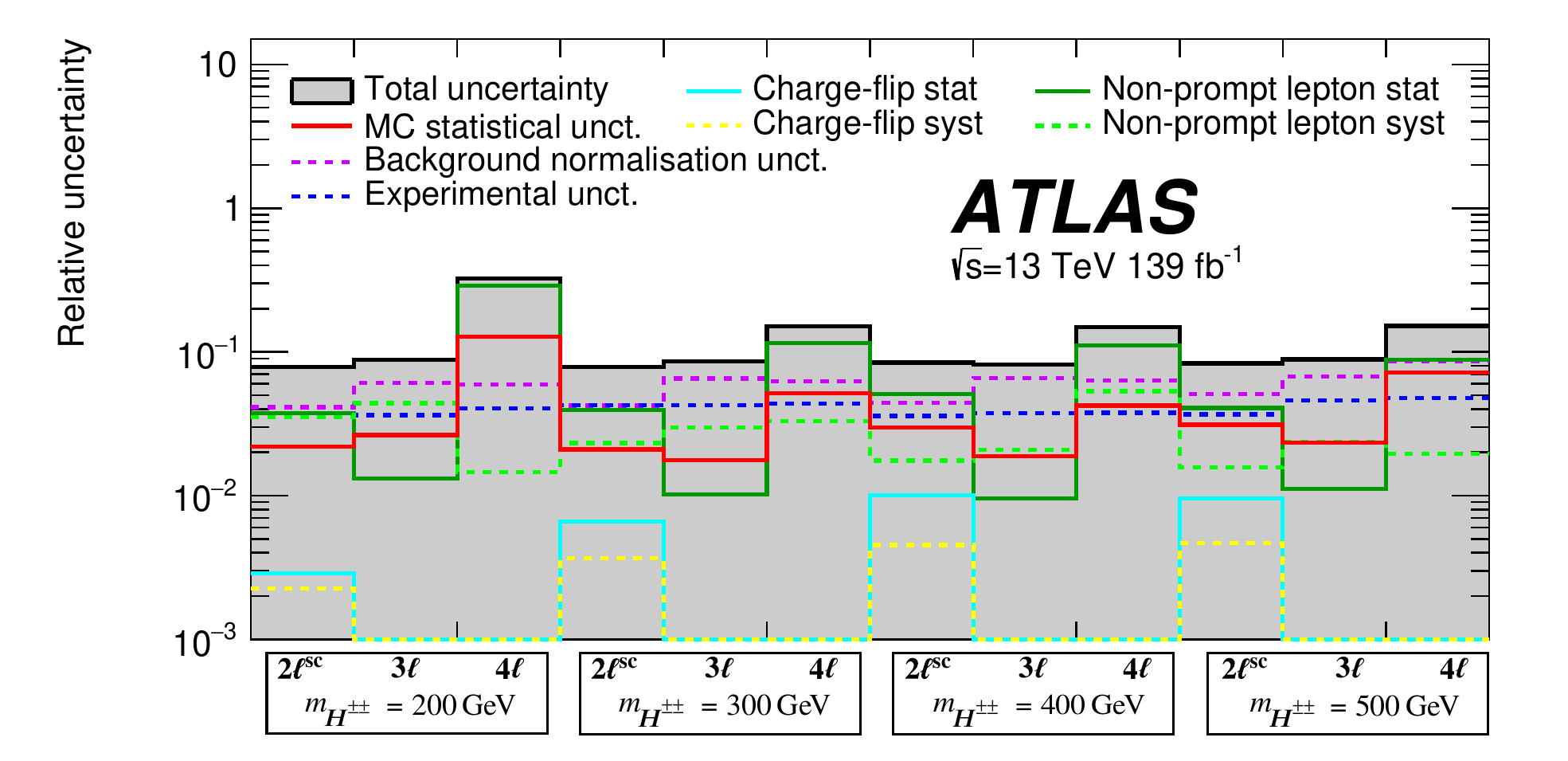}
\caption{
Contributions from different categories of uncertainties relative to the expected background yields in the defined SRs.
The uncertainties are shown for the combination of the individual channels of the \ll, \lll and \llll SRs.
The $m_{H^{\pm\pm}} = \SI{220}{GeV}$, \SI{350}{GeV} and \SI{450}{GeV} signal hypotheses use the signal regions defined for the
$m_{H^{\pm\pm}} = \SI{200}{GeV}$, \SI{300}{GeV} and \SI{400}{GeV} signal.
The $m_{H^{\pm\pm}}=\SI{550}{GeV}$ and $m_{H^{\pm\pm}}=\SI{600}{GeV}$ signal hypotheses use the signal regions defined for the $m_{H^{\pm\pm}} = \SI{500}{GeV}$ signal.
The SRs are indicated along the horizontal axis.
}
\label{fig,Syst_SRs}
\end{figure}

The relative uncertainty in the background yields obtained in a fit of background to the observed data is shown in Figure~\ref{fig,Syst_SRs} for all SRs.
The statistical uncertainties originate from the limited number of preselected and opposite-charge data events used
in the fake-factor method and the charge-flip electron background estimate, respectively, as well as the effect of the limited
number of simulated events for SM processes with prompt leptons.
The total uncertainty is computed using all sources of uncertainty, and they are treated as uncorrelated.
The uncertainties range from 10\% to 30\% and are dominated by the
statistical uncertainties in the non-prompt-lepton estimate and the theory uncertainties.
An exception is the \ll SR defined for $m_{H^{\pm\pm}} = \SI{300}{GeV}$,
where the uncertainties from most sources are of similar size.
The uncertainties associated with the charge-flip background are small in all \ll SRs.
In the \llll channel, the statistical uncertainty of the non-prompt leptons comes from the limited number of events in the $Z$+jets, $t \bar{t}$ and $tW$ MC simulation samples.

The theoretical uncertainties in the predicted signal yields arise from the parton shower model, missing higher-order corrections, and parton distribution functions.
The systematic uncertainty due to the parton shower model is evaluated by comparing event samples simulated by \PYTHIA~{8} using the A14 tune with samples from \herwig using
the \textsc{H7-ue-mmht} underlying-event tune~\cite{Bellm:2015jjp}, and is found to be less than 5\%.
The uncertainties due to the PDFs are found to be less than 5\%.
The uncertainty due to missing higher-order corrections is less than 10\%~\cite{Muhlleitner:2003me}.
When those uncertainties are combined in quadrature, an overall uncertainty in the signal yields of approximately 10\% is obtained for the signal normalisation.

\section{Results} 
\label{sec:result}
\begin{figure}[t!]
\centering
\includegraphics[width=0.8\textwidth]{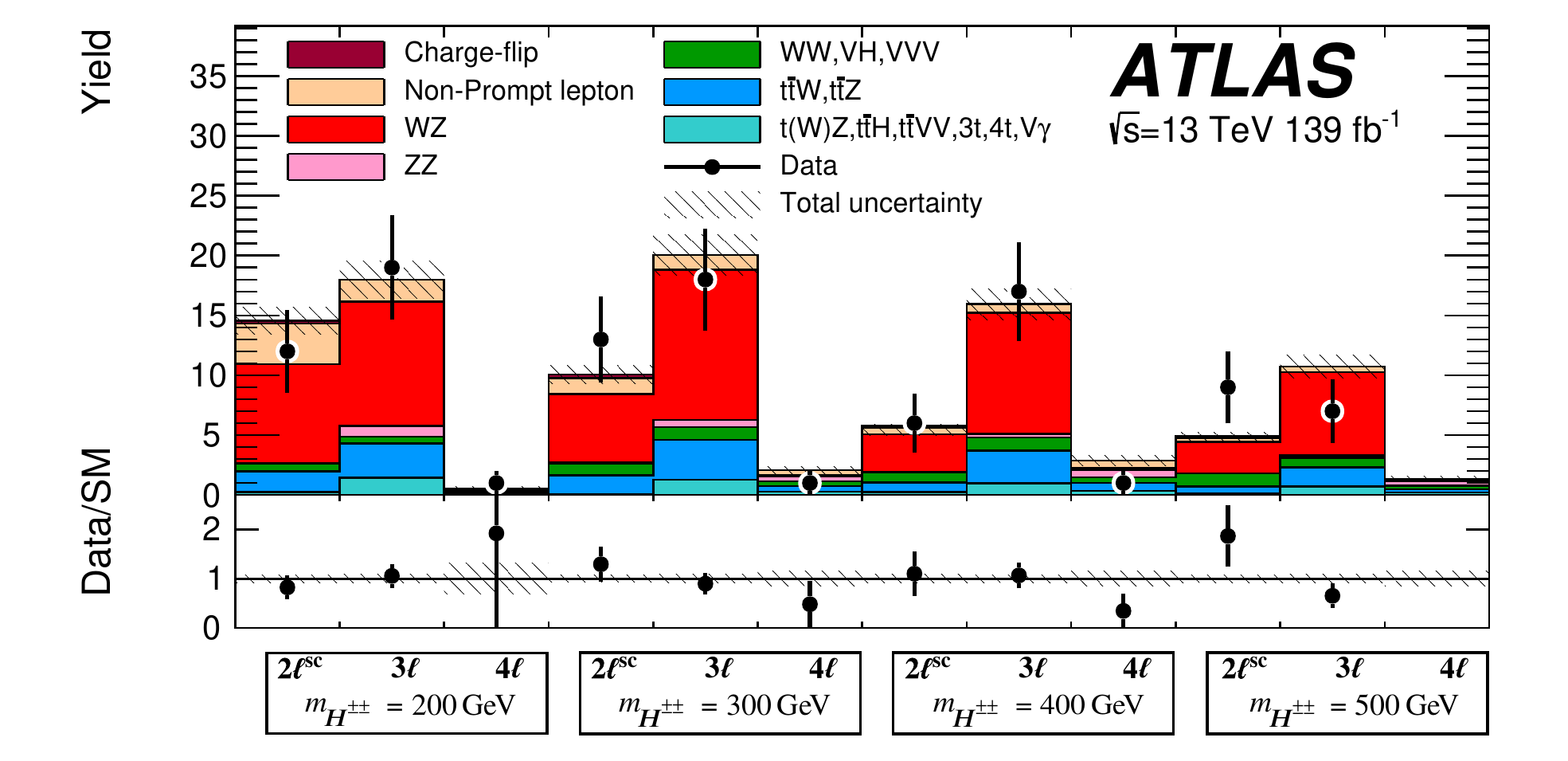}
\caption{
Data event yields compared with the expected contributions from relevant background sources,
for the combination of the individual channels of the \ll, \lll and \llll SRs.
The $m_{H^{\pm\pm}} = \SI{220}{GeV}$, \SI{350}{GeV} and \SI{450}{GeV} signal hypotheses use the signal regions defined for the
$m_{H^{\pm\pm}} = \SI{200}{GeV}$, \SI{300}{GeV} and \SI{400}{GeV} signal.
The $m_{H^{\pm\pm}}=\SI{550}{GeV}$ and $m_{H^{\pm\pm}}=\SI{600}{GeV}$ signal hypotheses use the signal regions defined for the $m_{H^{\pm\pm}} = \SI{500}{GeV}$ signal.
The total uncertainties in the expected event yields are shown as the hatched bands.
}
\label{fig,yields_SRs}
\end{figure}

\begin{table}[t!]
\begin{center}
\def\arraystretch{1.2}
\caption{
The expected background and the observed data event yields in the signal region defined for the $m_{H^{\pm\pm}} = \SI{300}{GeV}$ mass hypothesis.
The signal yield is for the corresponding mass point and is normalised to the luminosity of \SI{139}{fb^{-1}}.
The displayed numbers include all sources of statistical and systematic uncertainties.
The overall signal acceptances $A_\mathrm{PP}$ and $A_\mathrm{AP}$ and the observed upper limit on extra contributions to each signal region at 95\% confidence level, $n_{95}$, are also presented.
Selections with $\mu\mu$, three or four leptons are not affected by the electron charge-flip background, so these contributions are denoted by ``--''.
}
\label{Tab:EventYield}
 
\scriptsize
\resizebox{0.88\textwidth}{!}
{
\begin{tabular}{l||ccc|cc|c}
\hline\hline
\multirow{3}{*}{\hspace{3mm}SR}    & \multicolumn{3}{c|}{\ll} & \multicolumn{2}{c|}{\lll} & \multirow{3}{*}{\llll}                                                                                                          \\
\cline{2-6}
& \multirow{2}{*}{$ee$}    & \multirow{2}{*}{$e\mu$}   & \multirow{2}{*}{$\mu\mu$} & \multicolumn{2}{c|}{Number of same-flavour opposite-charge pairs} &                                 \\
&                          &                           &                           & 0                                                                 & $> 0$         &                 \\
\hline\hline

\hspace{3mm}Prompt lepton          & 1.66$\pm$0.28            & 4.3$\pm$0.5               & 2.30$\pm$0.26             & 1.62$\pm$0.20                                                     & 17.2$\pm$1.6  & 1.69$\pm$0.19   \\
\hspace{3mm}Charge-flip            & 0.17$\pm$0.07            & $\numRP{0.102}{2}$$\pm$$\numRP{0.034}{2}$   & --                        & --                                                                & --            & --              \\
\hspace{3mm}Non-prompt lepton      & 0.3$\pm$0.25             & 0.65$\pm$0.33             & 0.39$\pm$0.19             & 0.36$\pm$0.23                                                     & 0.9$\pm$0.6   & 0.41$\pm$0.25   \\
\hline
\hspace{3mm}Total background       & 2.1$\pm$0.4              & 5.1$\pm$0.6               & 2.69$\pm$0.32             & 1.98$\pm$0.29                                                     & 18.1$\pm$1.6  & 2.10$\pm$0.30   \\
\hspace{3mm}Data                   & 4                        & 8                         & 1                         & 1                                                                 & 17            & 1               \\
\hline
\hspace{3mm}$H^{\pm\pm}H^{\mp\mp}$ & 1.99$\pm$0.24            & 5.3$\pm$0.6               & 3.03$\pm$0.35             & 2.63$\pm$0.30                                                     & 7.6$\pm$0.9   & 1.50$\pm$0.17   \\
\hspace{3mm}$A_\mathrm{PP}$ [\%]   & $0.087$                  & $0.233$                   & $0.132$                   & $0.115$                                                           & $0.333$       & $0.065$         \\
\hline
\hspace{3mm}$H^{\pm\pm}H^{\mp}$    & 0.57$\pm$0.07            & 1.43$\pm$0.16             & 0.81$\pm$0.09             & 0.43$\pm$0.05                                                     & 1.35$\pm$0.16 & $\numRP{0.156}{2}$$\pm$$\numRP{0.020}{2}$ \\
\hspace{3mm}$A_\mathrm{AP}$ [\%]   & $0.043$                  & $0.109$                   & $0.062$                    & $0.033$                                                          & $0.103$       & $0.012$         \\
\hline
\hspace{3mm}$n_{95}$               & $6.72$                   & $9.21$                    & $3.24$                    & $3.27$                                                            & $9.52$        & $3.31$          \\

\hline\hline
\end{tabular}
}
\end{center}
\end{table}
 
 
The observed data event yields and the corresponding estimates for
the backgrounds in the SRs are shown in Figure~\ref{fig,yields_SRs}.
More details of the event yields in the signal region defined for $m_{H^{\pm\pm}} = \SI{300}{GeV}$ are given in Table~\ref{Tab:EventYield}.
No significant excess over the expected yields is observed in any of the SRs.
Table~\ref{Tab:EventYield} includes the acceptance for the signal from pair production of $H^{\pm\pm}$ bosons, $A_\mathrm{PP}$, and from associated production of $H^{\pm\pm}$ and $H^{\pm}$ bosons, $A_\mathrm{AP}$.
It is defined as the number of selected reconstructed events divided by the number of events at the event generation stage
and represents the signal reduction due to phase-space acceptance, branching ratio and detector efficiency.
The results in all the SRs are shown in the Appendix, in Table~\ref{Tab:EventYield_AllSRs}.

\begin{figure}[t!]
\centering
\subfloat[]{\includegraphics[width=0.335\textwidth]{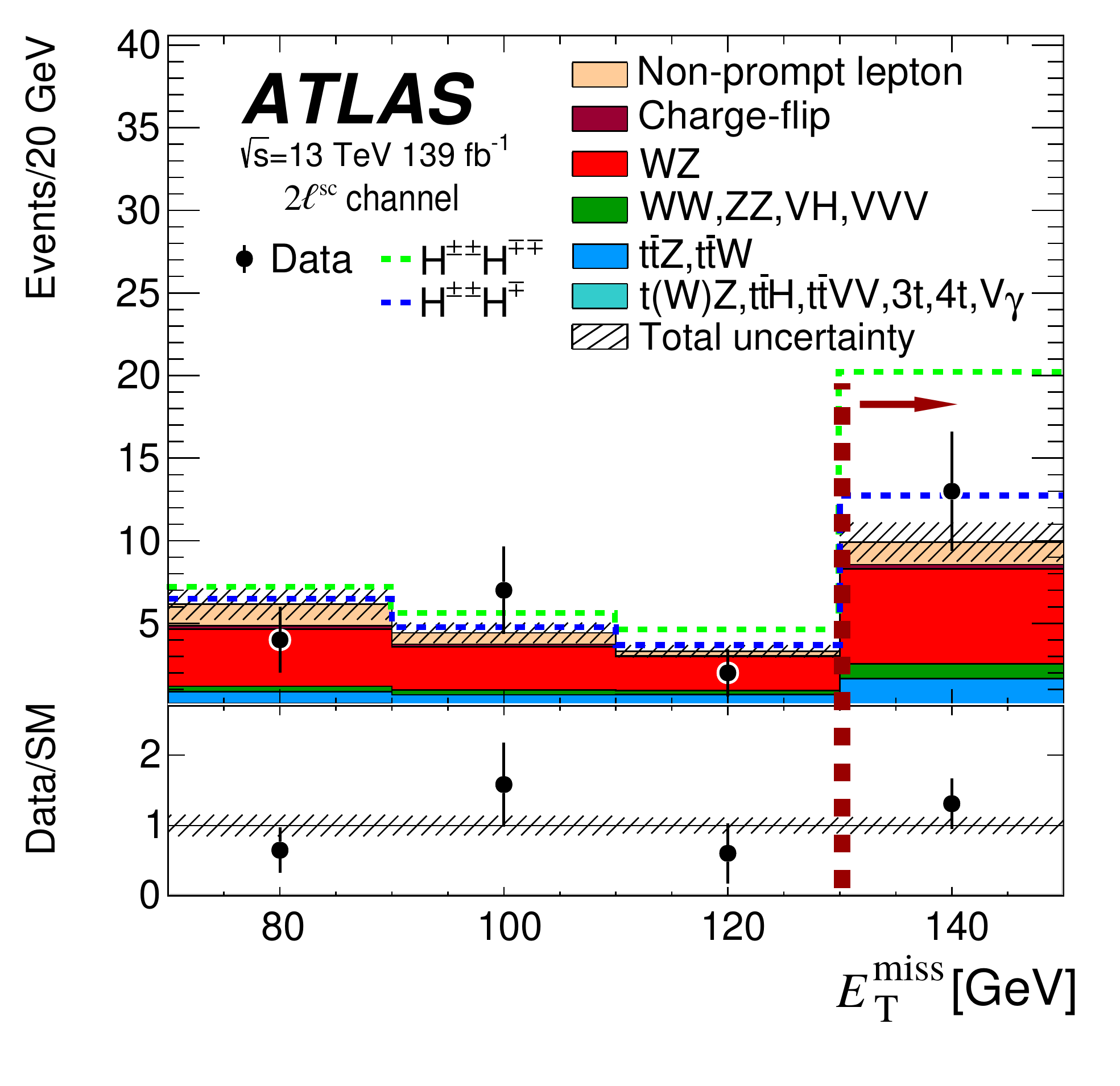}} \hspace{-0.2cm}
\subfloat[]{\includegraphics[width=0.335\textwidth]{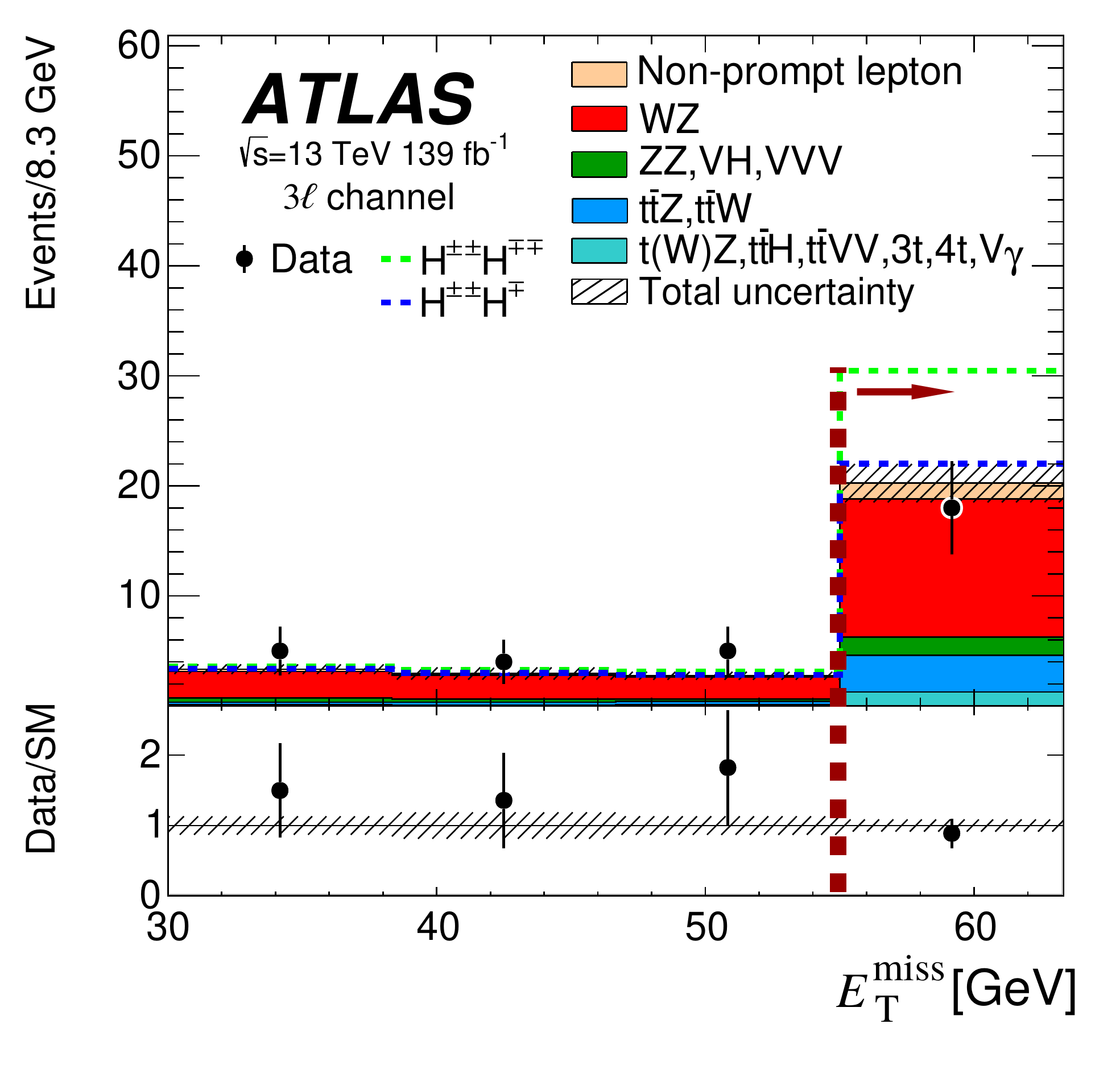}}\hspace{-0.2cm}
\subfloat[]{\includegraphics[width=0.335\textwidth]{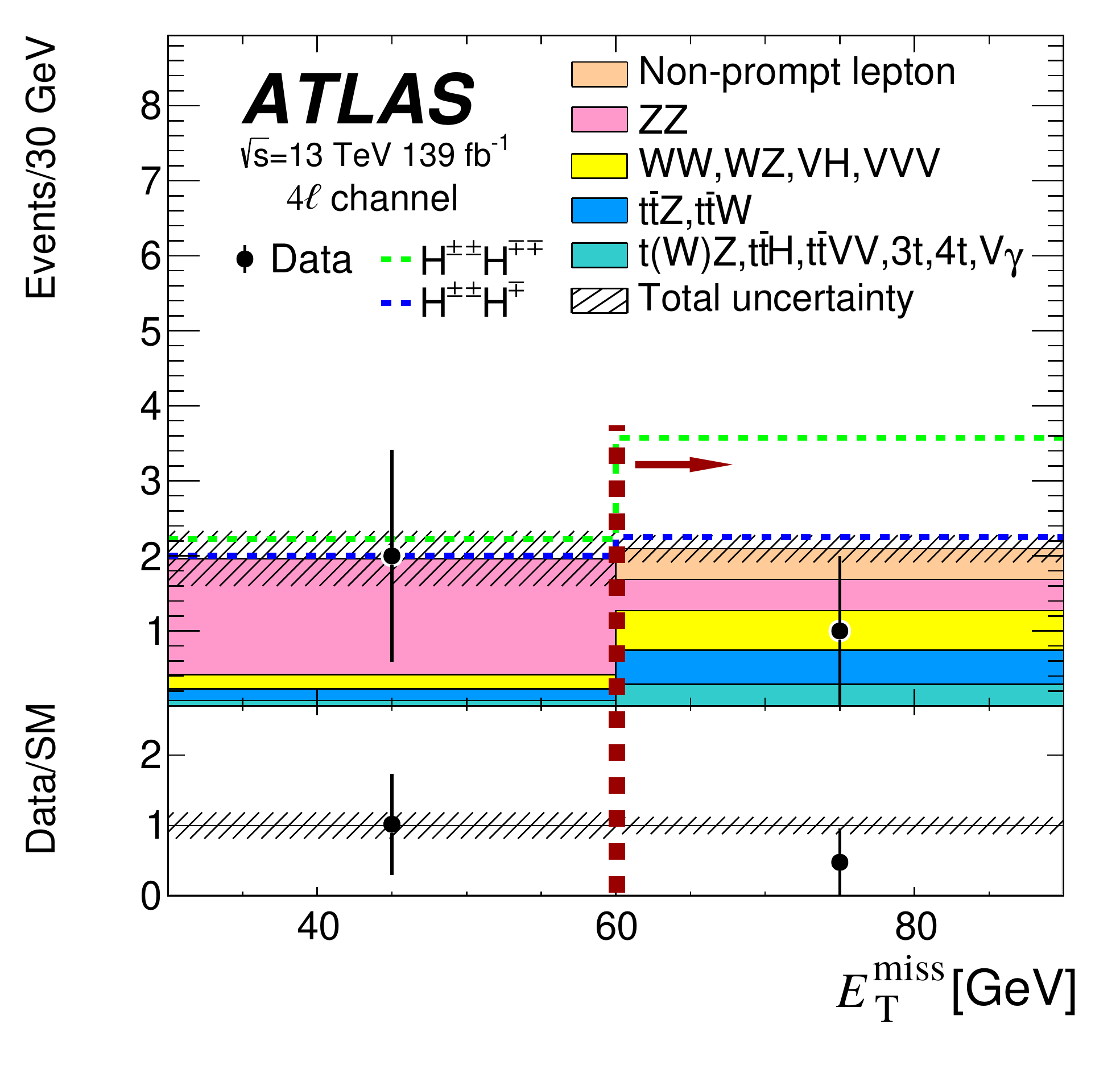}}
 
\caption{
The \met distribution for the SRs of the $m_{H^{\pm\pm}} = \SI{300}{GeV}$ signal mass hypothesis, where the selection requirement on \met has been removed.
The distribution is shown for the combination of the individual channels of the (a) \ll, (b) \lll and (c) \llll SRs.
The total uncertainties in the expected event yields are shown as hashed bands.
The last bin, isolated by a vertical red dashed line, is inclusive and corresponds to the SR.
The expected signal distributions corresponding to the $m_{H^{\pm\pm}} = \SI{300}{GeV}$ are also shown (stack histogram, on top of the background).
In each figure the bottom panel shows the ratio of data to the estimated background.
The hatched band around unity represents the total uncertainty of the background.
}
\label{fig,N-1 plots}
\end{figure}
 
The \met distribution is shown in Figure~\ref{fig,N-1 plots} for the SRs of the $m_{H^{\pm\pm}} = \SI{300}{GeV}$ signal mass hypothesis, where the selection requirement on \met has been removed.
The other criteria summarised in Table~\ref{Tab:SRselection} used to define the SR are applied.
Good agreement between data and the expected event yields is observed for low values of \met, demonstrating that the background contributions are also well estimated close to the SRs, where the
requirements on the kinematic variables are tighter than at preselection.

\begin{figure}[t!]
\subfloat[]{
\includegraphics[width=.48\textwidth]{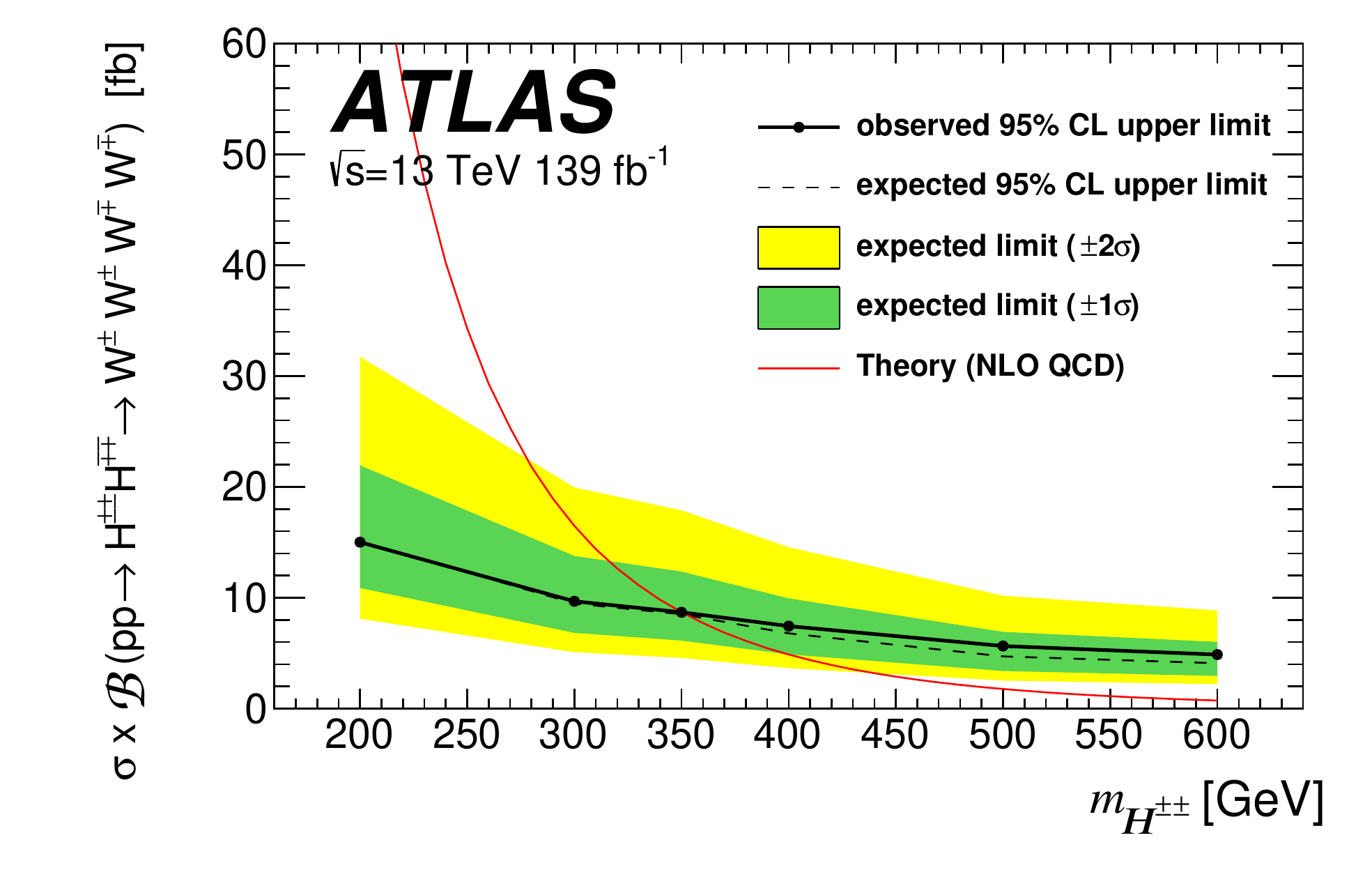}
\label{fig:ExpectedLimits_PP}
}
\subfloat[]{
\includegraphics[width=.48\textwidth]{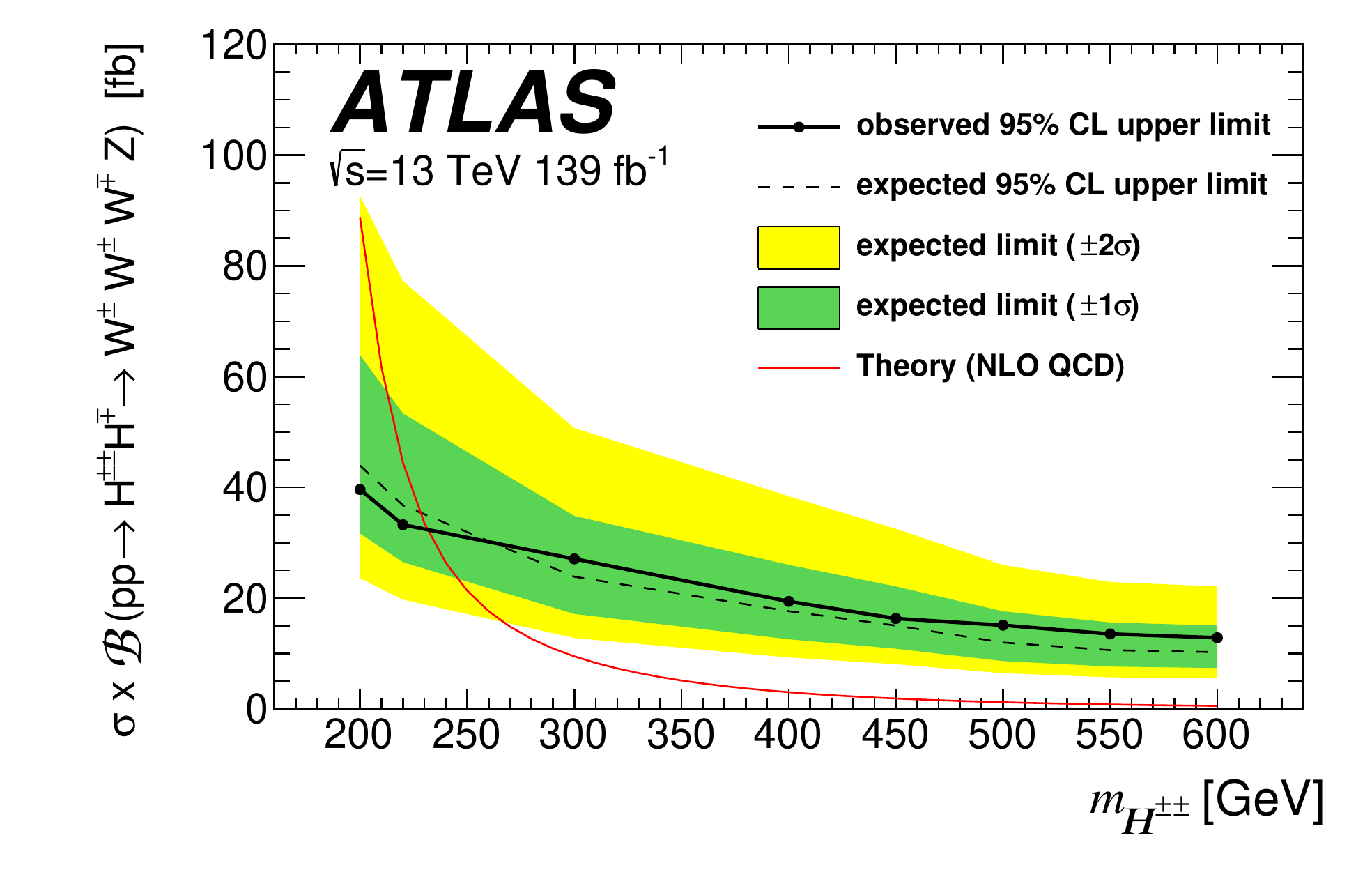}
\label{fig:ExpectedLimits_AP}
}
\caption{
Observed and expected upper limits on the
(a) $H^{\pm\pm}$ pair production and (b) $H^{\pm\pm}$ and $H^{\pm}$ associated production cross section times branching fraction
at $95\%$ CL obtained from the combination of \ll, \lll and \llll channels.
The region above the observed limit is excluded by the measurement.
The bands represent the expected exclusion curves within one and two standard deviations.
The theoretical prediction~\cite{PhysRevD.84.095005} including the NLO QCD corrections~\cite{Muhlleitner:2003me} is also shown.
}
\end{figure}

The statistical interpretation is based on a likelihood ratio test~\cite{CLREF} using the CL$_\mathrm{s}$ method~\cite{cls}. 
The signal strength, a free parameter in the fit, modifies the cross section of the signal hypothesis under investigation.
A separate  likelihood function is constructed for every signal hypothesis as the product of the Poisson probability distributions of the six channels of the corresponding signal region.
Gaussian distributions are used to constrain the nuisance parameters associated with the systematic uncertainties.
The widths of the Gaussian distributions correspond to the magnitudes of these uncertainties.
Statistical uncertainties in the backgrounds are estimated using Poisson distributions.

The expected and observed upper limits
on the $H^{\pm\pm}$ pair production and the $H^{\pm\pm}$ and $H^{\pm}$ associated production cross sections times branching fraction
at 95\% confidence level (CL) for the five mass hypotheses
are shown in Figures~\ref{fig:ExpectedLimits_PP} and~\ref{fig:ExpectedLimits_AP}, respectively.
They are obtained from the combination of \ll, \lll and \llll SRs.
Assuming a linear dependence of the cross-section limit between neighbouring mass points,
the observed 95\% CL lower limit on the mass of the $H^{\pm\pm}$ boson is $\SI{350}{GeV}$ for the pair production mode
and $\SI{230}{GeV}$ for the associated production mode.
To confirm the validity of the linear extrapolations close to the exclusion threshold,
upper limits on the $H^{\pm\pm}$ pair production cross section times branching fraction are also computed for the complementary mass hypothesis of $\SI{350}{GeV}$,
as shown in Figure~\ref{fig:ExpectedLimits_PP}, and are found to match the extrapolated values well.
A similar test was also done for the $H^{\pm\pm}$ and $H^{\pm}$ associated production mode,
and the upper limits on the cross section times branching fraction are computed for the \SI{220}{GeV}, \SI{450}{GeV} and \SI{550}{GeV} complementary mass hypotheses.
The results are shown in Figure~\ref{fig:ExpectedLimits_AP}, and again good matching is found.
 
A tighter limit on the $H^{\pm\pm}$ boson mass is obtained for the $H^{\pm\pm}$ pair production mode mainly due to the
cross sections times branching fraction for $H^{\pm\pm}$ pair production being higher than those for associated production of $H^{\pm\pm}$ and $H^{\pm}$.
Other important reasons are the different branching ratios of $H^{\pm\pm}H^{\mp}$ ($\approx 16\%$) and $H^{\pm\pm}H^{\mp\mp}$ ($\approx 26\%$) decays into \ll, \lll and \llll leptons,
and the signal acceptance, which is higher for the pair production mode than for the associated production mode (see Tables~\ref{Tab:EventYield} and~\ref{Tab:EventYield_AllSRs}).
The definition of common signal regions also plays a role.
As discussed in Section~\ref{sec:EvtSelection}, the SRs are not optimal for associated production of $H^{\pm\pm}$ and $H^{\pm}$.
For similar reasons, the upper limits on the charged Higgs boson production cross sections times branching fraction are weaker for associated production of $H^{\pm\pm}$ and $H^{\pm}$ than for $H^{\pm\pm}$ pair production.

For associated production of $H^{\pm\pm}$ and $H^{\pm}$, the limit on the $H^{\pm\pm}$ boson mass
implies a constraint on the $H^{\mp}$ boson mass which is at most \SI{5}{GeV} different from the $H^{\pm\pm}$ mass.
Upper limits on the $H^{\pm\pm}$ pair ($H^{\pm\pm}$ and $H^{\pm}$ associated) production cross section times branching fraction range from
\SI{15}{fb} to \SI{5}{fb} (\SI{40}{fb} to \SI{10}{fb}) when the assumed $H^{\pm\pm}$ mass varies from \SI{200}{GeV} to \SI{600}{GeV}.
Observed upper limits at 95\% CL on the number of BSM events for each signal region are derived using the CL$_\mathrm{s}$ prescription and the results are shown in Table~\ref{Tab:EventYield} for $m_{H^{\pm\pm}} = \SI{300}{GeV}$ and for all SRs in the Appendix, in Table~\ref{Tab:EventYield_AllSRs}.

\FloatBarrier
\section{Conclusion} 
\label{sec:conclusion}
A search for pair production of $H^{\pm\pm}$ bosons, and
for associated production of $H^{\pm\pm}$ and $H^{\pm}$ bosons, in the context of a type-II seesaw model, is presented.
For the $H^{\pm\pm}$ pair production mode the $H^{\pm\pm}$ boson mass is at least \SI{100}{GeV} lower than the $H^{\pm}$ boson mass,
while for the associated production of $H^{\pm\pm}$ and $H^{\pm}$ bosons the mass difference between the doubly and singly charged Higgs bosons is at most \SI{5}{GeV}.
Only the bosonic decays $H^{\pm\pm}\rightarrow W^\pm W^\pm$ and $H^{\pm}\rightarrow W^\pm Z$ are studied.
Dedicated signal regions in the \ll, \lll and \llll channels are defined as a function of different mass hypotheses within the model to look for evidence of $H^{\pm\pm}$ and $H^{\pm}$ bosons
in the $\sqrt{s}=\SI{13}{TeV}$ proton--proton collision data sample collected between 2015 and 2018 with the ATLAS detector at the LHC.
The data sample corresponds to an integrated luminosity of \SI{139}{fb^{-1}}.
The increase in the total integrated luminosity allowed significant improvements in the lepton fake factors measurement and the estimation of the associated uncertainties.
The data are found to be in good agreement with the estimated background for all channels investigated.
Combining those channels, the model considered is excluded at 95\% confidence level for
$H^{\pm\pm}$ boson masses below $\SI{350}{GeV}$ for the pair production of $H^{\pm\pm}$ bosons,
and below $\SI{230}{GeV}$ for the associated production of $H^{\pm\pm}$ and $H^{\pm}$ bosons.
Upper limits on the pair (associated) production cross section times branching fraction range
from \SI{15}{fb} to \SI{5}{fb} (\SI{40}{fb} to \SI{10}{fb}) when the assumed $H^{\pm\pm}$ mass varies from \SI{200}{GeV} to \SI{600}{GeV}.
The results obtained for the $H^{\pm\pm}$ pair production scenario raise the exclusion limits beyond those from
a previous, similar search by ATLAS using a smaller \SI{13}{TeV} data set by approximately \SI{130}{GeV}.
In addition, the associated production of $H^{\pm\pm}$ and $H^{\pm}$ bosons is also explored, extending the previous search.
 
\clearpage
\appendix
\part*{Appendix}
\addcontentsline{toc}{part}{Appendix}
 
The expected background and the observed data event yields in the signal regions are shown in Table~\ref{Tab:EventYield_AllSRs}.
The signal yield is for the corresponding mass point and is normalised to the luminosity of \SI{139}{fb^{-1}},
and the displayed numbers include all sources of statistical and systematic uncertainties.
The overall signal acceptances $A_\mathrm{PP}$ and $A_\mathrm{AP}$ and the observed upper limit on extra contributions to each signal region at 95\% confidence level, $n_{95}$, are also presented.
The data and background yields obtained for
$m_{H^{\pm\pm}}=\SI{200}{GeV}$, \SI{300}{GeV} and \SI{400}{GeV} are also valid for
$m_{H^{\pm\pm}}=\SI{220}{GeV}$, \SI{350}{GeV} and \SI{450}{GeV}, respectively.
The data and background yields obtained for $m_{H^{\pm\pm}}=\SI{500}{GeV}$ signal regions are also valid for $m_{H^{\pm\pm}}=\SI{550}{GeV}$ and $m_{H^{\pm\pm}}=\SI{600}{GeV}$.
Selections with $\mu\mu$, three or four leptons are not affected by the electron charge-flip background, so these contributions are denoted by ``--''.
 
\begin{table}[htbp!]
\begin{center}
\def\arraystretch{1.2}
\caption{
The expected background and the observed data event yields in the signal regions. More details are given in the text.
}
\label{Tab:EventYield_AllSRs}
 
\scriptsize
\resizebox{0.8\textwidth}{!}
{
\begin{tabular}{l||ccc|cc|c}
\hline\hline
\multirow{3}{*}{\hspace{3mm}SR}    & \multicolumn{3}{c|}{\ll} & \multicolumn{2}{c|}{\lll} & \multirow{3}{*}{\llll}                                                                                                               \\
\cline{2-6}
& \multirow{2}{*}{$ee$}    & \multirow{2}{*}{$e\mu$}   & \multirow{2}{*}{$\mu\mu$} & \multicolumn{2}{c|}{Number of same-flavour opposite-charge pairs} &                                      \\
&                          &                           &                           & 0                                                                 & $> 0$            &                   \\

\hline\hline
\multicolumn{7}{l}{$m_{H^{\pm\pm}} = \SI{200}{GeV}$}                                                                                                                                                                             \\
\hline\hline
 
\hspace{3mm}Prompt lepton          & 1.90$\pm$0.31            & 5.8$\pm$0.7               & 3.12$\pm$0.36             & 1.29$\pm$0.17                                                     & 14.9$\pm$1.4~~   & 0.36$\pm$0.07     \\
\hspace{3mm}Charge-flip            & 0.13$\pm$0.04            & $\numRP{0.118}{2}$$\pm$$\numRP{0.035}{2}$   & --                        & --                                                                & --               & -- \\
\hspace{3mm}Non-prompt lepton      & 1.0$\pm$0.5              & 1.45$\pm$0.5~~            & 0.93$\pm$0.31             & 0.7$\pm$0.4                                                       & 1.11$\pm$0.71    & 0.15$\pm$0.15     \\
\hspace{3mm}Total background       & 3.1$\pm$0.5              & 7.3$\pm$0.9               & 4.0$\pm$0.5               & 2.0$\pm$0.4                                                       & 15.97$\pm$1.48~~ & 0.51$\pm$0.17     \\
\hspace{3mm}Data                   & 3                        & 7                         & 2                         & 3                                                                 & 16               & 1                 \\
\hspace{3mm}$n_{95}$               & $4.95$                   & $6.68$                    & $3.78$                    & $5.43$                                                            & $9.78$           & $3.68$            \\
\hspace{3mm}$H^{\pm\pm}H^{\mp\mp}$ & 6.2$\pm$0.8              & 18.9$\pm$2.2~~            & 11.3$\pm$1.3~~            & 8.3$\pm$1.0                                                       & 23.16$\pm$2.68~~ & 3.29$\pm$0.38     \\
\hspace{3mm}$A_\mathrm{PP}$ [\%]   & $0.055$                  & $0.168$                   & $0.100$                   & $0.074$                                                           & $0.206$          & $0.029$           \\
\hspace{3mm}$H^{\pm\pm}H^{\mp}$    & 2.8$\pm$0.4              & 9.0$\pm$1.1               & 5.1$\pm$0.6               & 2.32$\pm$0.29                                                     & 6.56$\pm$0.76    & 0.36$\pm$0.06     \\
\hspace{3mm}$A_\mathrm{AP}$ [\%]   & $0.022$                  & $0.073$                   & $0.042$                   & $0.019$                                                           & $0.053$          & $0.003$           \\
 
\hline\hline
\multicolumn{7}{l}{$m_{H^{\pm\pm}} = \SI{220}{GeV}$}                                                                                                                                                                             \\
\hline\hline
\hspace{3mm}$H^{\pm\pm}H^{\mp}$    & 1.74$\pm$0.21            & 5.1$\pm$0.6               & 3.1$\pm$0.4               & 1.45$\pm$0.18                                                     & 4.1$\pm$0.5      & 0.23$\pm$0.04     \\
\hspace{3mm}$A_\mathrm{AP}$ [\%]   & $0.028$                  & $0.082$                   & $0.050$                   & $0.023$                                                           & $0.066$          & $0.004$           \\
\hline\hline
\multicolumn{7}{l}{$m_{H^{\pm\pm}} = \SI{300}{GeV}$}                                                                                                                                                                             \\
\hline\hline
 
\hspace{3mm}Prompt lepton          & 1.66$\pm$0.28            & 4.3$\pm$0.5               & 2.30$\pm$0.26             & 1.62$\pm$0.20                                                     & 17.2$\pm$1.6~~   & 1.69$\pm$0.19     \\
\hspace{3mm}Charge-flip            & 0.17$\pm$0.07            & $\numRP{0.102}{2}$$\pm$$\numRP{0.034}{2}$  & --                    & --                                                                & --               & --   \\
\hspace{3mm}Non-prompt lepton      & ~~0.3$\pm$0.25           & 0.65$\pm$0.33             & 0.39$\pm$0.19             & 0.36$\pm$0.23                                                     & 0.9$\pm$0.6      & 0.41$\pm$0.25     \\
\hspace{3mm}Total background       & 2.1$\pm$0.4              & 5.1$\pm$0.6               & 2.69$\pm$0.32             & 1.98$\pm$0.29                                                     & 18.1$\pm$1.6~~   & 2.10$\pm$0.30     \\
\hspace{3mm}Data                   & 4                        & 8                         & 1                         & 1                                                                 & 17               & 1                 \\
\hspace{3mm}$n_{95}$               & $6.72$                   & $9.21$                    & $3.24$                    & $3.27$                                                            & $9.52$           & $3.31$            \\
\hspace{3mm}$H^{\pm\pm}H^{\mp\mp}$ & 1.99$\pm$0.24            & 5.3$\pm$0.6               & 3.03$\pm$0.35             & 2.63$\pm$0.30                                                     & 7.6$\pm$0.9      & 1.50$\pm$0.17     \\
\hspace{3mm}$A_\mathrm{PP}$ [\%]   & $0.087$                  & $0.233$                   & $0.132$                   & $0.115$                                                           & $0.333$          & $0.065$           \\
\hspace{3mm}$H^{\pm\pm}H^{\mp}$    & 0.57$\pm$0.07            & 1.43$\pm$0.16             & 0.81$\pm$0.09             & 0.43$\pm$0.05                                                     & 1.35$\pm$0.16    & $\numRP{0.156}{2}$$\pm$$\numRP{0.020}{2}$   \\
\hspace{3mm}$A_\mathrm{AP}$ [\%]   & $0.043$                  & $0.109$                   & $0.062$                   & $0.033$                                                           & $0.103$          & $0.012$           \\

\hline\hline
\multicolumn{7}{l}{$m_{H^{\pm\pm}} = \SI{350}{GeV}$}                                                                                                                                                                             \\
\hline\hline
 
\hspace{3mm}$H^{\pm\pm}H^{\mp\mp}$ & 1.20$\pm$0.14            & 3.11$\pm$0.35             & 1.74$\pm$0.20             & 1.57$\pm$0.18                                                     & 4.5$\pm$0.5      & 0.89$\pm$0.10     \\
\hspace{3mm}$A_\mathrm{PP}$ [\%]   & $0.099$                  & $0.257$                   & $0.144$                   & $0.130$                                                           & $0.372$          & $0.074$           \\
 
\hline\hline
\multicolumn{7}{l}{$m_{H^{\pm\pm}} = \SI{400}{GeV}$}                                                                                                                                                                             \\
\hline\hline
 
\hspace{3mm}Prompt lepton          & 0.87$\pm$0.12            & 2.52$\pm$0.29             & 1.33$\pm$0.16             & 1.41$\pm$0.18                                                     & 13.8$\pm$1.2~~   & 2.26$\pm$0.25     \\
\hspace{3mm}Charge-flip            & 0.09$\pm$0.04            & 0.10$\pm$0.04             & --                        & --                                                                & --               & --                \\
\hspace{3mm}Non-prompt lepton      & 0.12$\pm$0.17            & 0.24$\pm$0.20             & 0.19$\pm$0.13             & 0.22$\pm$0.15                                                     & 0.5$\pm$0.33     & 0.6$\pm$0.4       \\
\hspace{3mm}Total background       & 1.08$\pm$0.21            & 2.86$\pm$0.35             & 1.52$\pm$0.21             & 1.63$\pm$0.22                                                     & 14.3$\pm$1.2~~   & 2.9$\pm$0.4       \\
\hspace{3mm}Data                   & 3                        & 3                         & 0                         & 3                                                                 & 14               & 1                 \\
\hspace{3mm}$n_{95}$               & $6.25$                   & $5.03$                    & $2.48$                    & $5.73$                                                            & $9.05$           & $3.33$            \\
\hspace{3mm}$H^{\pm\pm}H^{\mp\mp}$ & 0.66$\pm$0.08            & 1.73$\pm$0.19             & ~~1.0$\pm$0.11            & 0.91$\pm$0.11                                                     & 2.65$\pm$0.30    & 0.63$\pm$0.07     \\
\hspace{3mm}$A_\mathrm{PP}$ [\%]   & $0.097$                  & $0.255$                   & $0.147$                   & $0.134$                                                           & $0.391$          & $0.093$           \\
\hspace{3mm}$H^{\pm\pm}H^{\mp}$    & $\numRP{0.186}{2}$$\pm$$\numRP{0.023}{2}$ & 0.48$\pm$0.06             & $\numRP{0.271}{2}$$\pm$$\numRP{0.032}{2}$ & $\numRP{0.156}{2}$$\pm$$\numRP{0.019}{2}$ & 0.48$\pm$0.06    & $\numRP{0.107}{2}$$\pm$$\numRP{0.013}{2}$   \\
\hspace{3mm}$A_\mathrm{AP}$ [\%]   & $0.046$                  & $0.115$                   & $0.065$                   & $0.038$                                                           & $0.115$          & $0.026$           \\
 
\hline\hline
\multicolumn{7}{l}{$m_{H^{\pm\pm}} = \SI{450}{GeV}$}                                                                                                                                                                             \\
\hline\hline
\hspace{3mm}$H^{\pm\pm}H^{\mp}$    & $\numRP{0.126}{2}$$\pm$$\numRP{0.016}{2}$ & 0.34$\pm$0.04             & $\numRP{0.200}{2}$$\pm$$\numRP{0.023}{2}$ & $\numRP{0.119}{2}$$\pm$$\numRP{0.015}{2}$& 0.36$\pm$0.04    & $\numRP{0.074}{2}$$\pm$$\numRP{0.010}{2}$   \\
\hspace{3mm}$A_\mathrm{AP}$ [\%]   & $0.048$                  & $0.129$                   & $0.076$                   & $0.045$                                                           & $0.136$          & $0.028$           \\
 
\hline\hline
\multicolumn{7}{l}{$m_{H^{\pm\pm}} = \SI{500}{GeV}$}                                                                                                                                                                             \\
\hline\hline
 
\hspace{3mm}Prompt lepton          & 0.97$\pm$0.14            & 2.25$\pm$0.27             & 1.12$\pm$0.15             & 0.92$\pm$0.14                                                     & 9.4$\pm$0.9      & 1.23$\pm$0.17     \\
\hspace{3mm}Charge-flip            & $\numRP{0.089}{2}$$\pm$$\numRP{0.032}{2}$   & 0.08$\pm$0.04             & --                        & --                                                                & --               & --                \\
\hspace{3mm}Non-prompt lepton      & 0.23$\pm$0.19            & $<0.1$                    & 0.10$\pm$0.09             & 0.06$\pm$0.06                                                     & 0.4$\pm$0.27     & 0.13$\pm$0.12     \\
\hspace{3mm}Total background       & 1.29$\pm$0.23            & 2.33$\pm$0.27             & 1.22$\pm$0.17             & 0.98$\pm$0.14                                                     & 9.8$\pm$0.9      & 1.36$\pm$0.20     \\
\hspace{3mm}Data                   & 4                        & 4                         & 1                         & 1                                                                 & 6                & 0                 \\
\hspace{3mm}$n_{95}$               & $7.63$                   & $6.54$                    & $3.46$                    & $3.56$                                                            & $5.13$           & $2.47$            \\
\hspace{3mm}$H^{\pm\pm}H^{\mp\mp}$ & 0.37$\pm$0.04            & 0.85$\pm$0.1              & 0.45$\pm$0.05             & 0.39$\pm$0.04                                                     & 1.12$\pm$0.13    & $\numRP{0.243}{2}$$\pm$$\numRP{0.028}{2}$   \\
\hspace{3mm}$A_\mathrm{PP}$ [\%]   & $0.150$                  & $0.346$                   & $0.183$                   & $0.159$                                                           & $0.455$          & $0.098$           \\
\hspace{3mm}$H^{\pm\pm}H^{\mp}$    & $\numRP{0.114}{2}$$\pm$$\numRP{0.014}{2}$            & $\numRP{0.263}{2}$$\pm$$\numRP{0.030}{2}$       & $\numRP{0.140}{2}$$\pm$$\numRP{0.017}{2}$  & $\numRP{0.074}{2}$$\pm$$\numRP{0.009}{2}$ & $\numRP{0.242}{2}$$\pm$$\numRP{0.028}{2}$  & $\numRP{0.051}{2}$$\pm$$\numRP{0.006}{2}$   \\
\hspace{3mm}$A_\mathrm{AP}$ [\%]   & $0.068$                  & $0.158$                   & $0.084$                   & $0.044$                                                           & $0.145$          & $0.031$           \\
 
\hline\hline
\multicolumn{7}{l}{$m_{H^{\pm\pm}} = \SI{550}{GeV}$}                                                                                                                                                                             \\
\hline\hline
\hspace{3mm}$H^{\pm\pm}H^{\mp}$    & $\numRP{0.089}{2}$$\pm$$\numRP{0.011}{2}$   & $\numRP{0.195}{2}$$\pm$$\numRP{0.023}{2}$ & $\numRP{0.105}{2}$$\pm$$\numRP{0.012}{2}$ & $\numRP{0.053}{2}$$\pm$$\numRP{0.006}{2}$ & $\numRP{0.174}{2}$$\pm$$\numRP{0.020}{2}$  & $\numRP{0.039}{2}$$\pm$$\numRP{0.005}{2}$   \\
\hspace{3mm}$A_\mathrm{AP}$ [\%]   & $0.08$                   & $0.175$                   & $0.094$                   & $0.048$    & $0.156$          & $0.035$           \\
\hline\hline
\multicolumn{7}{l}{$m_{H^{\pm\pm}} = \SI{600}{GeV}$}                                                                                                                                                                             \\
\hline\hline
 
\hspace{3mm}$H^{\pm\pm}H^{\mp\mp}$ & $\numRP{0.170}{2}$$\pm$$\numRP{0.020}{2}$  & 0.40$\pm$0.05 & $\numRP{0.220}{2}$$\pm$$\numRP{0.025}{2}$ & $\numRP{0.182}{2}$$\pm$$\numRP{0.021}{2}$ & 0.54$\pm$0.06    & $\numRP{0.112}{2}$$\pm$$\numRP{0.013}{2}$   \\
\hspace{3mm}$A_\mathrm{PP}$ [\%]   & $0.167$                  & $0.394$                   & $0.217$                   & $0.177$                                                           & $0.532$          & $0.108$           \\
\hspace{3mm}$H^{\pm\pm}H^{\mp}$    & $\numRP{0.058}{2}$$\pm$$\numRP{0.007}{2}$   & $\numRP{0.133}{2}$$\pm$$\numRP{0.015}{2}$ & $\numRP{0.070}{2}$$\pm$$\numRP{0.008}{2}$     & $\numRP{0.038}{2}$$\pm$$\numRP{0.005}{2}$ & $\numRP{0.125}{2}$$\pm$$\numRP{0.015}{2}$  & $\numRP{0.0277}{2}$$\pm$$\numRP{0.0034}{2}$ \\
\hspace{3mm}$A_\mathrm{AP}$ [\%]   & $0.083$                  & $0.191$                   & $0.101$                   & $0.055$                                                           & $0.180$          & $0.040$           \\

\hline\hline
\end{tabular}
}
\end{center}
\end{table}

\clearpage

\section*{Acknowledgements}

We thank CERN for the very successful operation of the LHC, as well as the
support staff from our institutions without whom ATLAS could not be
operated efficiently.
 
We acknowledge the support of ANPCyT, Argentina; YerPhI, Armenia; ARC, Australia; BMWFW and FWF, Austria; ANAS, Azerbaijan; SSTC, Belarus; CNPq and FAPESP, Brazil; NSERC, NRC and CFI, Canada; CERN; ANID, Chile; CAS, MOST and NSFC, China; COLCIENCIAS, Colombia; MSMT CR, MPO CR and VSC CR, Czech Republic; DNRF and DNSRC, Denmark; IN2P3-CNRS and CEA-DRF/IRFU, France; SRNSFG, Georgia; BMBF, HGF and MPG, Germany; GSRT, Greece; RGC and Hong Kong SAR, China; ISF and Benoziyo Center, Israel; INFN, Italy; MEXT and JSPS, Japan; CNRST, Morocco; NWO, Netherlands; RCN, Norway; MNiSW and NCN, Poland; FCT, Portugal; MNE/IFA, Romania; JINR; MES of Russia and NRC KI, Russian Federation; MESTD, Serbia; MSSR, Slovakia; ARRS and MIZ\v{S}, Slovenia; DST/NRF, South Africa; MICINN, Spain; SRC and Wallenberg Foundation, Sweden; SERI, SNSF and Cantons of Bern and Geneva, Switzerland; MOST, Taiwan; TAEK, Turkey; STFC, United Kingdom; DOE and NSF, United States of America. In addition, individual groups and members have received support from BCKDF, CANARIE, Compute Canada, CRC and IVADO, Canada; Beijing Municipal Science \& Technology Commission, China; COST, ERC, ERDF, Horizon 2020 and Marie Sk{\l}odowska-Curie Actions, European Union; Investissements d'Avenir Labex, Investissements d'Avenir Idex and ANR, France; DFG and AvH Foundation, Germany; Herakleitos, Thales and Aristeia programmes co-financed by EU-ESF and the Greek NSRF, Greece; BSF-NSF and GIF, Israel; La Caixa Banking Foundation, CERCA Programme Generalitat de Catalunya and PROMETEO and GenT Programmes Generalitat Valenciana, Spain; G\"{o}ran Gustafssons Stiftelse, Sweden; The Royal Society and Leverhulme Trust, United Kingdom.
 
The crucial computing support from all WLCG partners is acknowledged gratefully, in particular from CERN, the ATLAS Tier-1 facilities at TRIUMF (Canada), NDGF (Denmark, Norway, Sweden), CC-IN2P3 (France), KIT/GridKA (Germany), INFN-CNAF (Italy), NL-T1 (Netherlands), PIC (Spain), ASGC (Taiwan), RAL (UK) and BNL (USA), the Tier-2 facilities worldwide and large non-WLCG resource providers. Major contributors of computing resources are listed in Ref.~\cite{ATL-SOFT-PUB-2020-001}.
 

\clearpage
\printbibliography
\clearpage
\input{atlas_authlist}

\end{document}

%% file: atlas_authlist.tex
 
\begin{flushleft}
\hypersetup{urlcolor=black}
{\Large The ATLAS Collaboration}

\bigskip

\AtlasOrcid[0000-0002-6665-4934]{G.~Aad}$^\textrm{\scriptsize 101}$,    
\AtlasOrcid[0000-0002-5888-2734]{B.~Abbott}$^\textrm{\scriptsize 127}$,    
\AtlasOrcid[0000-0002-7248-3203]{D.C.~Abbott}$^\textrm{\scriptsize 102}$,    
\AtlasOrcid[0000-0002-2788-3822]{A.~Abed~Abud}$^\textrm{\scriptsize 36}$,    
\AtlasOrcid[0000-0002-1002-1652]{K.~Abeling}$^\textrm{\scriptsize 53}$,    
\AtlasOrcid[0000-0002-2987-4006]{D.K.~Abhayasinghe}$^\textrm{\scriptsize 93}$,    
\AtlasOrcid[0000-0002-8496-9294]{S.H.~Abidi}$^\textrm{\scriptsize 29}$,    
\AtlasOrcid[0000-0002-8279-9324]{O.S.~AbouZeid}$^\textrm{\scriptsize 40}$,    
\AtlasOrcid{N.L.~Abraham}$^\textrm{\scriptsize 155}$,    
\AtlasOrcid[0000-0001-5329-6640]{H.~Abramowicz}$^\textrm{\scriptsize 160}$
% ...
\bigskip
\\

% ...
$^{29}$Physics Department, Brookhaven National Laboratory, Upton NY; United States of America.\\
% ...
$^{36}$CERN, Geneva; Switzerland.\\
% ...
$^{40}$Niels Bohr Institute, University of Copenhagen, Copenhagen; Denmark.\\
% ...
$^{53}$II. Physikalisches Institut, Georg-August-Universit\"{a}t G\"ottingen, G\"ottingen; Germany.\\
% ...
$^{93}$Department of Physics, Royal Holloway University of London, Egham; United Kingdom.\\
% ...
$^{101}$CPPM, Aix-Marseille Universit\'e, CNRS/IN2P3, Marseille; France.\\
$^{102}$Department of Physics, University of Massachusetts, Amherst MA; United States of America.\\
% ...
$^{127}$Homer L. Dodge Department of Physics and Astronomy, University of Oklahoma, Norman OK; United States of America.\\
% ...
$^{155}$Department of Physics and Astronomy, University of Sussex, Brighton; United Kingdom.\\
% ...
$^{160}$Raymond and Beverly Sackler School of Physics and Astronomy, Tel Aviv University, Tel Aviv; Israel.\\
% ...

\end{flushleft}
